\documentclass[acmsmall]{acmart}

\AtBeginDocument{%
  \providecommand\BibTeX{{%
    \normalfont B\kern-0.5em{\scshape i\kern-0.25em b}\kern-0.8em\TeX}}}

\setcopyright{acmcopyright}
\acmJournal{PACMHCI}
\acmYear{2019}
\acmVolume{3}
\acmNumber{CSCW}
\acmArticle{183}
\acmMonth{11}
\acmPrice{15.00}
\acmDOI{10.1145/3359285}

\usepackage{enumitem}
\usepackage{subfig}
\usepackage[nameinlink,capitalize]{cleveref}

\usepackage{xspace}
\newcommand{\vmg}{\textsc{Vevo Music Graph}\xspace}
\newcommand{\vevo}{\textsc{Vevo}\xspace}
\newcommand{\header}[1]{{\noindent{\textbf{#1}}}}

\usepackage{xcolor}
\usepackage{soul}
\definecolor{navy}{rgb}{0.0, 0.0, 0.0}
\definecolor{ruby}{rgb}{0.0, 0.0, 0.0}
\definecolor{ao}{rgb}{0.0, 0.0, 0.0}

\begin{document}

%
\title{Estimating Attention Flow in Online Video Networks}

%
\author{Siqi Wu}
\affiliation{%
  \institution{Australian National University \& Data 61, CSIRO}
  \country{Australia}
}
\email{siqi.wu@anu.edu.au}

\author{Marian-Andrei Rizoiu}
\affiliation{%
  \institution{University of Technology Sydney \& Data 61, CSIRO}
  \country{Australia}
}
\email{marian-andrei.rizoiu@uts.edu.au }

\author{Lexing Xie}
\affiliation{%
  \institution{Australian National University \& Data 61, CSIRO}
  \country{Australia}
}
\email{lexing.xie@anu.edu.au}

%
\renewcommand{\shortauthors}{Siqi Wu, Marian-Andrei Rizoiu, \& Lexing Xie}

%

\begin{abstract}
Online videos have shown tremendous increase in Internet traffic.
Most video hosting sites implement recommender systems, which connect the videos into a directed network and conceptually act as a source of pathways for users to navigate.
At present, little is known about how human attention is allocated over such large-scale networks, and about the impacts of the recommender systems.
In this paper, we first construct the \vevo network --- a YouTube video network with 60,740 music videos interconnected by the recommendation links, and we collect their associated viewing dynamics.
This results in a total of 310 million views every day over a period of 9 weeks.
Next, we present large-scale measurements that connect the structure of the recommendation network and the video attention dynamics.
We use the bow-tie structure to characterize the \vevo network and we find that its core component (23.1\% of the videos), which occupies most of the attention (82.6\% of the views), is made out of videos that are mainly recommended among themselves.
This is indicative of the links between video recommendation and the inequality of attention allocation.
Finally, we address the task of estimating the attention flow in the video recommendation network.
We propose a model that accounts for the network effects for predicting video popularity, and we show it consistently outperforms the baselines.
This model also identifies a group of artists gaining attention because of the recommendation network.
Altogether, our observations and our models provide a new set of tools to better understand the impacts of recommender systems on collective social attention.
\end{abstract}

%
%


%

%
\maketitle


\section{Introduction}
\label{sec:intro}


Many online platforms present algorithmic suggestions to help users explore the enormous content space.
The recommender systems, which produce such suggestions, are central to modern online platforms.
They have been employed in many applications, such as finding new friends on Twitter~\cite{su2016effect}, discovering interesting communities on LinkedIn~\cite{sharma2013pairwise}, and recommending similar goods on Amazon~\cite{oestreicher2012recommendation,dhar2014prediction}.
In the domain of multimedia, service providers (e.g., YouTube, Netflix, and Spotify) use recommender systems to suggest related videos or songs~\cite{davidson2010youtube,covington2016deep,gomez2016netflix,zhang2012auralist,celma2008hits}.
Much effort has been on generating more accurate recommendations, but relatively little is said about the effects of recommender systems on overall attention, such as their effects on item popularity ranking, the estimated strength of item-to-item links, and global patterns on the attention gained due to being recommended.
This work aims to answer such questions for online videos, using publicly available recommendation networks and attention time series. 

We use the term \textit{attention} to refer to a broad range of user activities with respect to an online item, such as clicks, views, likes, comments, shares, or time spent watching. 
The term \textit{popularity}, however, is used to denote observed attention statistics that are often used to rank online items against each other. 
In this work, our measurement and estimation are carried out on the largest online video platform YouTube (as of 2019), and we specifically quantify popularity using the number of daily views for each video.
The outlined methods may well apply to other deeper forms of user engagement such as watch time.
Due to data availability constraints, the validation in this work is limited to popularity. 


\begin{figure*}[tb]
	\centering
	\subfloat{\includegraphics[width=1\textwidth]{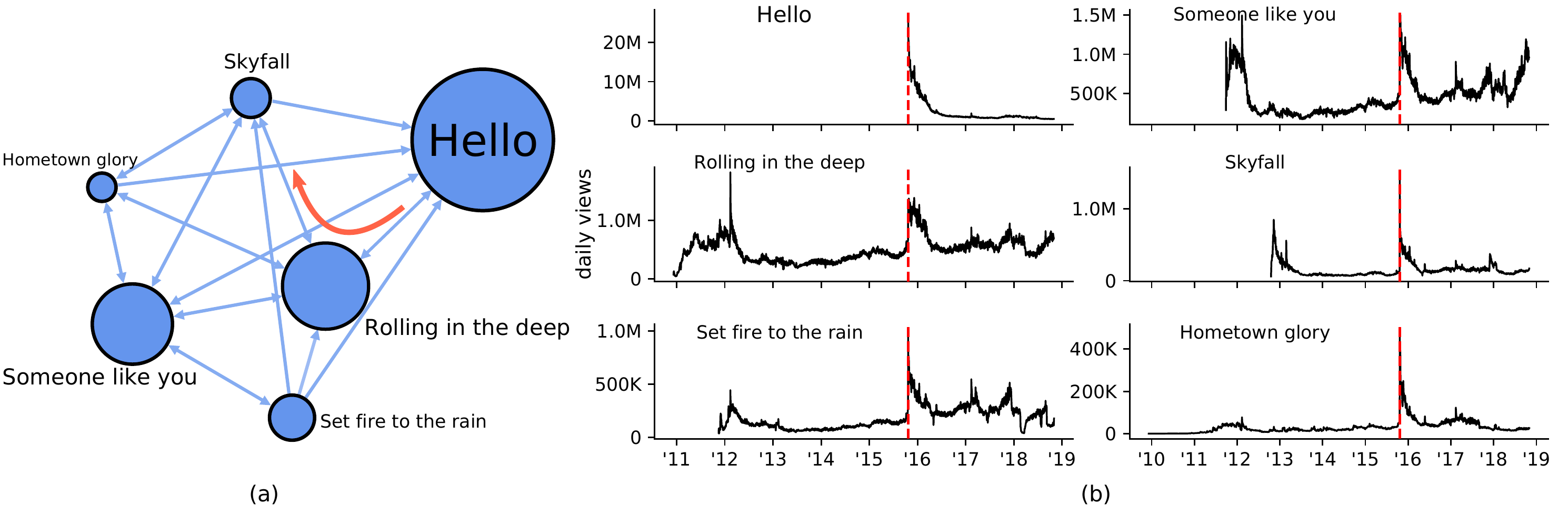}}
	\caption{Observing the effects of recommendation network on video popularity.
	\textbf{(a)} A directed network consists of six videos by the artist Adele.
	The node size is proportional to the video's cumulative view counts till Nov 02, 2018.
	The red arrow highlights one possible route that users visit ``Skyfall'' from ``Hello'' in 2 hops.
	\textbf{(b)} View series for the six videos shown in \textbf{(a)}. Visually we observe a simultaneous spike across all videos when ``Hello'' was uploaded on Oct 22, 2015, denoted by red dashed vertical line.
	}
	\label{fig:teasers}
\end{figure*}

We illustrate the goals of this work through an example.
\cref{fig:teasers}(a) shows the recommendation network for six videos from the artist Adele.
It is a directed network and the directions imply how users can navigate between videos by following the recommendation links.
Some videos are not directly connected but reachable within a few hops.
For example, ``Skyfall'' is not on the recommended list of ``Hello'', but a user can visit ``Skyfall'' from ``Hello'' by first visiting ``Rolling in the deep''.
\cref{fig:teasers}(b) plots the daily view series since the upload of each of the six videos.
When ``Hello'' was released, it broke the YouTube debut records by attracting 28M views in the first 24 hours~\cite{billboard2015adele}.
Simultaneously, we observe a traffic spike in all of her other videos, even in three videos that were not directly pointed by ``Hello''.
This example illustrates that the viewing dynamics of videos connected directly or indirectly through recommendation links may correlate, and it prompts us to investigate the patterns of attention flowing between them.

This work bridges two gaps in the current literature.
The first gap measures and estimates the effects of recommender systems in complex social systems.
The main goals of recommender systems are maximizing the chance that a user clicks on an item in the next step~\cite{davidson2010youtube,covington2016deep,bendersky2014up,yi2014beyond} or in a longer time horizon~\cite{beutel2018latent,chen2019top,ie2019slateq}.
However, recommendation in social systems remains as an open problem for two reasons:
(1) a limited conceptual understanding of how finite human attention is allocated over the network of content, in which some items gain popularity at the expense of, or with the assistance of others;
(2) the computational challenge of jointly recommending a large collection of items.
The second gap comes from a lack of fine-grained measurements on the attention captured by items structured as a network.
There are recent measurements on the YouTube recommendation networks~\cite{airoldi2016follow,cheng2008statistics}, but their measurements are not connected to the attention patterns on content.
Similarly, measurement studies on YouTube attention~\cite{zhou2010impact} quantify the overall volume of views directed from recommended links.
However, no measurement that accounts for both the network structure and the attention flow is available for online videos.

This paper tackles three research questions:

\begin{enumerate}[label=\textbf{RQ\arabic*:}]
  \item How to measure video recommendation network from publicly available information?
  \item What are the characteristics of the video recommendation network?
  \item Can we estimate the attention flow in the video recommendation network?
\end{enumerate}

We address the first question by curating a new YouTube dataset consisting of a large set of \vevo artists.
This is the first dataset that records both the temporal network snapshots of a recommender system, and the attention dynamics for items in it.
Our observation window lasts 9 weeks.
We present two means to construct the non-personalized recommendation network, and we discuss the relation between them in detail (\cref{sec:data}).

Addressing the second question, we conceptualize the global structure of the network as a bow-tie~\cite{broder2000graph} and we find that the largest strongly connected component accounts for $23.11\%$ of the videos while occupying $82.6\%$ of the attention.
Surprisingly, videos with high indegree are mostly songs with sustained interests, but not the latest released songs with high view counts.
We further find that the network structure is temporally consistent on the macroscopic level, however, there is a significant link turnover on the microscopic level.
For example, $50\%$ of the videos with an indegree of 100 on a particular day will gain or lose at least 10 links on the next day, and 25\% links appear only once during our 9-week observation window (\cref{sec:measures}).

Answering the third question, we build a model which employs both the temporal and network features to predict video popularity, and we estimate the amount of views flowing over each link.
Our networked model consistently outperforms the autoregressive and neural network baseline methods.
For an average video in our dataset, we estimate that $31.4\%$ of its views are contributed by the recommendation network.
We also find the evidence of YouTube recommender system boosting the popularity of some niche artists (\cref{sec:models}).

The new methods and observations in this work can be used by content owners, hosting sites, and online users alike.
For content owners, the understanding of how much traffic is driven among their own content or from/to other content can lead to better production and promotion strategies.
For hosting sites, such understanding can help avoid social optimization, and shed light on building a fair and transparent content recommender systems.
For online users, understanding how human attention is shaped by the algorithmic recommendation can help them be conscious of the relevance, novelty and diversity trade-offs in the content they are recommended to.

The main contributions of this work include:
\begin{itemize}[leftmargin=*]
  \item We curate a new YouTube dataset, called \vmg dataset\footnote{The code and datasets are publicly available at \url{https://github.com/avalanchesiqi/networked-popularity}}, which contains the daily snapshots of the video recommendation network over a span of 9 weeks, and the associated daily view series for each video since upload.
  \item We perform, to our knowledge, the first large-scale measurement study that connects the structure of the recommendation network with video attention dynamics.
  \item We propose an effective model that accounts for the network structure to predict video popularity and to estimate the attention flow over each recommendation link.
\end{itemize}



\section{Related work}
\label{sec:related}

In this section, we discuss three lines of research: design of (video) recommender systems, measurements on recommender systems, and studies on user attention towards online items.

\subsection{Recommender systems and video recommendation}
The goals of recommender systems can be summarized as two related yet distinct tasks.
The first task is user-centric, i.e., given users' profiles and past activities, finding a collection of items that might interest them~\cite{konstan2012recommender,covington2016deep}.
The resulting recommendations, often shown in user homepage feed, can be regarded as the entry point for the user action sequence.
The second task is item-centric, i.e., given the currently visited item, finding a ranked list of relevant items~\cite{davidson2010youtube,zhang2012auralist,gomez2016netflix}.
This can be regarded as recommending the next item in a sequence of actions.
In the same vein, we conceptualize and explain the behaviors on YouTube --- users start the action sequences by latent interests, and their subsequent actions are driven by network effects (see \cref{ssec:setting}).

\header{Recommender systems on YouTube.}
Recommender systems, along with YouTube search, have been shown as the two dominant factors driving user attention on YouTube~\cite{zhou2010impact}.
In 2010, \citet{davidson2010youtube} reported the usage of a collaborative filtering method in the YouTube recommender systems, i.e., videos are recommended by counting the number of co-watches.
This approach works well for videos with many views, however, it is less applicable for newly uploaded videos or least watched videos.
\citet{bendersky2014up} proposed two methods to enhance the collaborative filtering approach by embedding the video topic representation into the recommender.
\citet{covington2016deep} applied deep neural networks and indicated that the final recommendation is a top-K sample from a large candidate set generated by taking into the account content relevance, past watch and search activities, etc.
Other enhancements include incorporating contextual data \cite{beutel2018latent}.
Most recently, \citet{chen2019top} and \citet{ie2019slateq} showed success in applying reinforcement learning techniques in YouTube recommender systems.

Our work does not deal with designing a recommender system, nor does it attempt to reverse engineer the YouTube recommender.
Instead, we concentrate our analysis on the impacts of the recommender systems by presenting large-scale measurements.

\subsection{Measuring the effects of recommender systems}
Contrasting the extensive literature on evaluating the accuracy of recommendation~\cite{zhang2012auralist,beutel2018latent,chen2019top,li2018offline}, we focus on prior work that connects network structure with content consumption.
\citet{carmi2017oprah} reported how the book sales on Amazon react to exogenous demand shocks --- not only did the sales increase for the featured item, but the increase also
propagated a few hops away by following the links created by the recommender systems.
This is akin to our observation in \cref{fig:teasers} that attention ripples happen for videos too.
\citet{dhar2014prediction} further showed the effectiveness of using the recommendation network in predicting item demands.
\citet{su2016effect} linked the aggregate effects of recommendations and network structure, and found that popular items profit substantially more than the average ones.
However, \citet{sharma2015estimating} stressed the difficulty of inferring causal relations based on observational data in recommender systems.

\citet{cheng2008statistics} are among the first to study the statistics of YouTube recommender systems.
They scraped video webpages to construct the video network at a weekly interval.
\citet{airoldi2016follow} followed the video suggestions on YouTube to construct one static network snapshot for a random collection of music videos.
Note that both studies adopt a snowball sampling technique to construct the network, whereas in our work, we have the complete trace of an easily identifiable group of \vevo artists, and we capture the dynamics of network snapshots at a much finer daily granularity (see \cref{ssec:crawling}).
Most importantly, our work links the network with the item attention dynamics.

\subsection{Measuring and predicting online attention}
Attention is a scarce resource in online platforms.
While users have an unprecedented volume of information to choose from, online content competes for our limited attention~\cite{weng2012competition,zarezade2017correlated}.
\citet{salganik2006experimental} designed the ``MusicLab'' experiment, in which they explored how social influence and inherent quality affect a product's market share.
In a follow-up study, \citet{krumme2012quantifying} conceptualized user behaviors as two steps for characterizing how users consume digital items.
The first step is based on the appeal of the product, measured by the number of clicks; the second step is based on the quality of the product, measured by post-clicking metrics, e.g., dwell time, comments or shares.
A similar two-step process is employed in the web search community to differentiate between page views and dwell time on webpages~\cite{yue2010beyond,yi2014beyond}.
Following a similar idea, we categorize online attention into \textit{popularity} and \textit{engagement}.
On YouTube, popularity refers to the number of views that a video receives and engagement refers to the time spent on watching the video.

Predicting content popularity is an active field.
For online videos, future popularity has been shown to correlate with popularity in the past~\cite{pinto2013using,szabo2010predicting}, rendering autoregressive method a strong baseline.
Other works integrate additional information.
External sharing on social media has been linked to the popularity of online videos~\cite{li2013popularity,abisheva2014watches}, which is later developed by \citet{rizoiu2017expecting} to model popularity as an interplay of exogenous stimuli and endogenous responses.
Another line of work measures the temporal characteristics of content popularity.
\citet{yu2015lifecyle} revealed that the lifecycles of online videos exhibit a multi-phase pattern.
\citet{figueiredo2016trendlearner} stressed the necessity of predicting content popularity before the user interests exhaust.

For engagement studies on YouTube, we refer to our previous paper~\cite{wu2018beyond}, in which we found the most engaging videos do not always have highest view counts.
In contrast to the general understanding that content popularity is unpredictable in social systems~\cite{martin2016exploring,cheng2014can,rizoiu2018sir}, we also found that engagement metrics appear to be much more predictable.

This work focuses on the popularity measures.
To our knowledge, no prior work has attempted to predict video popularity with fine-grained recommendation network information due to the difficulty in constructing such network.
This is the first study that shows how to construct a persistent network by following the recommended links, and how to employ network features to improve the popularity prediction task (see \cref{sec:models}).
It is worth differentiating our work from the studies that collect individual user data from customized browser plugins.
Instead of measuring proactive user behaviors, we are interested in understanding how the platform-generated recommendation network guide the aggregate user attention.



\section{Constructing YouTube video network}
\label{sec:data}

In this section, we first introduce our newly curated \vmg dataset (\cref{ssec:vmg}).
Next, we detail the data collection strategy (\cref{ssec:crawling}) and analyze the relation between two types of non-personalized video recommendation lists (\cref{ssec:nonpersonal}).

\subsection{\vmg dataset}
\label{ssec:vmg}

The \vmg dataset consists of the verified \vevo artists who are active in six English-speaking countries (United States, United Kingdom, Canada, Australia, New Zealand, and Ireland), together with their complete record of videos uploaded on YouTube from the launch of \vevo (Dec 8, 2009) until Aug 31, 2018.
Our dataset contains 4,435 \vevo artists and 60,740 music videos.
For each video, we collect its metadata (e.g., title, description, uploader), its view count time series, and its recommendation relations with other videos.
The videos and their recommendation relations form a dynamic directed network, which we capture daily between Sep 1, 2018 and Nov 2, 2018 (63 days, 9 weeks).

\header{Why \vevo?}
\vevo\footnote{The \vevo website was shut down on May 24, 2018; however, videos syndicated on YouTube before are still embedded with a ``VEVO'' watermark on their thumbnails. See screenshot in \cref{fig:data-layout} for illustration.} is the largest syndication hub that provides licensed music videos from major record companies to YouTube \cite{wikipediavevo}.
We choose to study the networked attention flow on \vevo for several reasons.
First, \vevo is an ecosystem of its own that attracts tremendous attention --- 94 of all-time top 100 most viewed videos on YouTube are music, and 64 of which are distributed via \vevo \cite{wikipediatop}.
On average, our dataset accounts for 310 millions views and 9.1 millions watch hours every day.
Second, many users utilize YouTube as their music streaming player, listening to non-stop playlists generated by the recommender systems.
After the completion of the current video, YouTube automatically plays the ``Up next'' video --- the video in the first position of the recommended list, as illustrated in \cref{fig:data-layout}.
This usage pattern for music videos makes the network effects of YouTube recommender systems more significant for directing user attention from one video to another.
Third, \vevo artists and their videos form a tightly connected network.
The average degree in the \vevo video network is 10, compared to 3.2 in the YouTube video network collected by \citet{airoldi2016follow} via snowball sampling (see \cref{ssec:nonpersonal}).
The nodes are homogeneous in terms of content --- they are all music videos from artists based in English-speaking countries.
Lastly, the \vevo artists are easily identifiable --- they include the keyword ``VEVO'' in the channel title, they possess a verification badge on the channel page, and they publish licensed videos with a ``VEVO'' watermark.

\subsection{Data collection strategy}
\label{ssec:crawling}
We identify \vevo artists starting from Twitter. 
We capture every tweet that mentions YouTube videos by feeding the rule \texttt{"youtube" OR ("youtu" AND "be")} into the Twitter Streaming API\footnote{\url{https://developer.twitter.com/en/docs/tweets/filter-realtime/api-reference/post-statuses-filter.html}}.
Our Twitter crawler has been running continuously since Jun 2014.
From the ``\texttt{extended\_urls}'' field of each tweet, we extract the associated YouTube video ID, and we use our open-source tool \textsc{youtube-insight}~\cite{wu2018beyond} to retrieve the video's metadata, daily view count series and the ranked list of relevant videos.
Next, we select the \vevo artists by keeping only the channels that have the keyword ``VEVO'' in the channel title and a ``verified'' status badge on the channel homepage.
Note that a channel refers to a user who uploads videos on YouTube.
We query an open music database MusicBrainz\footnote{\url{https://musicbrainz.org}} to retrieve more features about each artist, such as the music genres and the geographical area of activities.
We retain the artists who are active in the six aforementioned English-speaking countries, and the videos that are classified into the ``Music'' category.
For completeness, we also implement a snowball-like procedure to retrieve further artists and their videos by following the recommendation relations from the tweeted videos.
However, this procedure only adds 2 more artists (out of the 4,435 \vevo artists in our dataset) and 5 more videos (out of the 60,740 music videos).
This is not surprising, considering most artists would promote their works on social media platforms.
One data limitation is that artists who are not affiliated with \vevo will not appear in our collection, such as Ed Sheeran and Christina Perry.

\subsection{The network of YouTube videos}
\label{ssec:nonpersonal}

\begin{figure*}[tbp]
	\centering
	\subfloat{\includegraphics[width=1\textwidth]{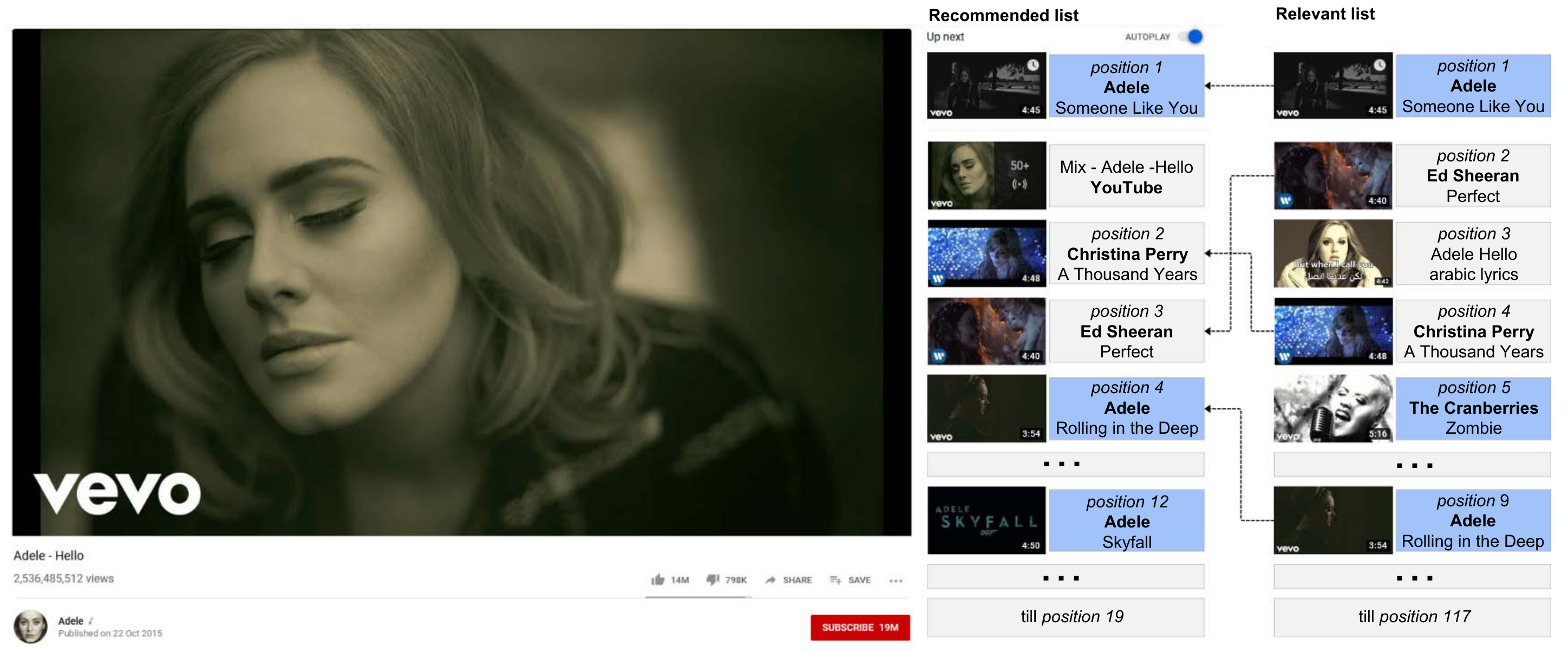}}
	\caption{
		YouTube webpage layout for the video ``Adele - Hello'', together with its recommended and relevant lists.
		\vevo videos are colored in blue and included in our dataset.
	}
	\label{fig:data-layout}
\end{figure*}

For any YouTube video, there are two publicly accessible sources of recommendation relations.
The first is the right-hand panel of videos that YouTube displays on its interface.
We denote this as the \textit{recommended list} (visualized in \cref{fig:data-layout}).
The second is from the YouTube Data API\footnote{\url{https://developers.google.com/youtube/v3/docs/search/list}}, which retrieves a list of videos that are relevant to the query video, ranked by the relevance.
We denote this as the \textit{relevant list}.
We retrieve both the recommended and the relevant lists for every video in our dataset.
We construct the recommended list by simulating a browser to access the video webpage and scraping the list on the right-hand panel.
We retrieve the first 20 videos from the panel, which are the default number of videos shown to the viewers on YouTube.
Note that typically, YouTube customizes the viewers' recommendation panel based on their personal interests and prior interaction history.
Here, we retrieve the non-personalized recommended list by sending all requests from a non-logged in client and by clearing the cookies before each request.
We denote the networks of videos constructed using the recommended and the relevant lists as the \textit{recommended network} and the \textit{relevant network}, respectively.

From Sep 1, 2018 to Nov 2, 2018, we crawled both the recommended and the relevant lists for each of the 60,740 \vevo videos on a daily basis.
The crawling jobs were distributed across 20 virtual machines, and took about 2 hours to finish.
In this way, we obtain successive snapshots for both the recommended and the relevant networks over 9 weeks.


\header{An illustrative example.}
\cref{fig:data-layout} illustrates the YouTube webpage layout for the video ``Hello'' by Adele, together with its recommended and relevant lists.
Videos belonging to the \vevo artists are colored in blue (e.g., Adele and The Cranberries), while others are colored in grey (e.g., Ed Sheeran and Christina Perry).
Visibly, not all videos on the recommended and relevant lists belong to the \vevo artists (e.g., ``Ed Sheeran - Perfect'').
Notice that for Music videos, a platform-generated playlist is always shown at the second position of the recommended list (here, ``Mix - Adele - Hello''), effectively capping the size of this list at 19.
The length of the relevant list often exceeds 100.
We observe that not all relevant videos appear in the recommended list (e.g., ``The Cranberries - Zombie''), nor all recommended videos originate from the relevant list (e.g., ``Adele - Skyfall'').
Also, the relative positions of two videos can appear flipped between the two lists (e.g., ``Ed Sheeran - Perfect'' and ``Christina Perry - A Thousand Years'').

\begin{figure*}[tbp]
	\centering
	\subfloat{\includegraphics[width=1\textwidth]{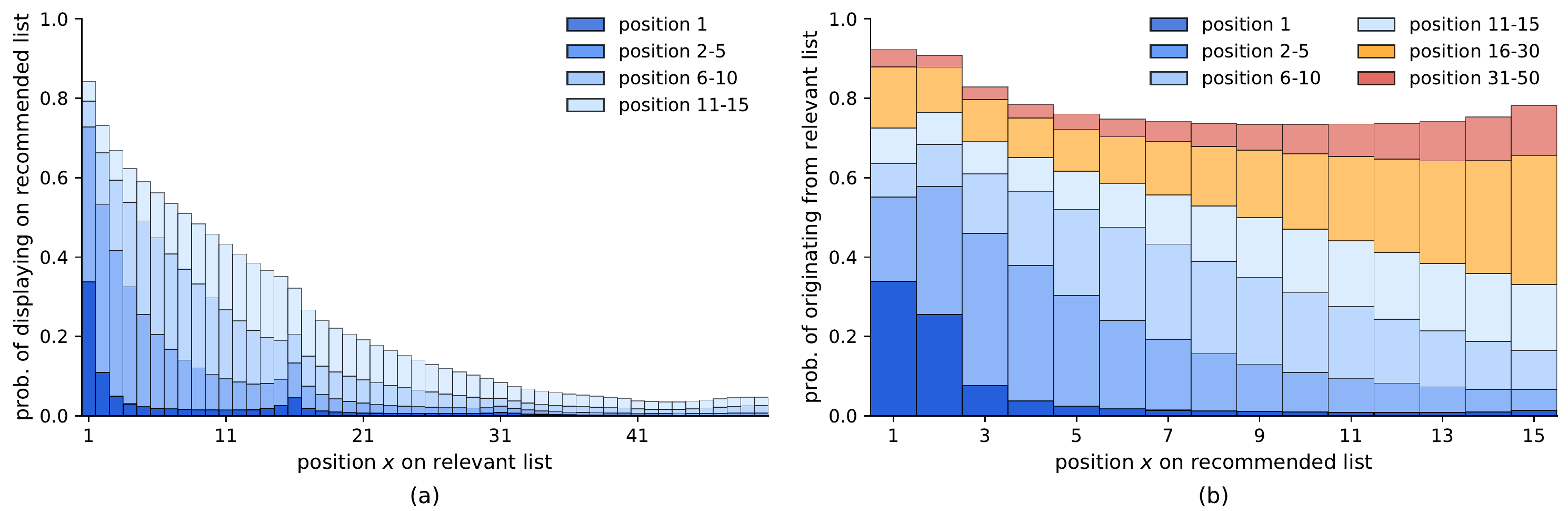}}
	\caption{ 
	\textbf{(a)} Display probabilities of videos from the position $x$ of the relevant list appear at different positions of the recommended list. 
	\textbf{(b)} Probabilities of videos from the position $x$ of the recommended list originate from different positions of the relevant list. 
	\textbf{(a)} and \textbf{(b)} are similar but with the relevant and recommended lists along the x-axis and y-axis swapped.
	}
	\label{fig:data-rel2rec}
\end{figure*}

\header{Display probabilities from the relevant to the recommended list.}
We study the relation between the positions of videos on the relevant and on the recommended lists.
We construct four bins based on the video position on the recommended list (position 1, position 2-5, position 6-10, and position 11-15).
\cref{fig:data-rel2rec}(a) shows as stacked bars the probability that a video ends up in each of the bins, as a function of its position on the relevant list.
The total height of the stacked bars gives the overall probability that a video originating from the relevant list appears at the top 15 positions on the recommended list.
We observe videos that appear at an upper position on the relevant list are more likely to appear on the recommended list, and at an upper position.
For example, the video at position 1 on the relevant list has 0.34 probability to appear at the first position and 0.84 probability to appear at the top 15 positions on the recommended list.
The probability decays for videos that appear at lower positions.
A relevant video appearing in position 41 to 50 has less than 0.05 probability to appear on the recommended list.
We compute the probabilities of appearance between each pair of positions in the relevant and the recommended lists --- denoted as \textit{display probabilities} --- using the 9-week dynamic network snapshots.


In \cref{fig:data-rel2rec}(b), we show the plot of the probability that a video at a given position on the recommended list originates from the relevant list.
We observe videos that appear at an upper position on the recommended list are more likely to originate from the upper position of relevant list.
The other notable observation is that the overall recall for recommended videos are high at over 0.8, meaning for any video on the recommended list, we are likely to see it on the relevant list.

\header{YouTube video network density.}
\citet{airoldi2016follow} used the first 25 videos on the relevant list to construct the relevant network, which had an average degree of $3.2$.
By comparison, our \vevo video network is much denser at the same cutoff, with an average degree of 10.
One could expect that the relevant network becomes even denser when videos at lower position are included; however, the display probabilities also need to be considered.
In this paper and unless otherwise specified, we use the first 15 positions on the relevant list ($0.35 \leq P_{\text{display}} \leq 0.84$) to construct the relevant network.
We denote this threshold as the \textit{cutoff} and we study the impact of different cutoff values on the network structure in \cref{sssec:bowtie}.
Measurements with other cutoff values yield similar results and thus are omitted.

\header{Discussion on the recommended and relevant lists.}
The notions of recommended and relevant lists have been previously adopted in the field
of recommender systems \cite{herlocker2004evaluating}.
The relevant list is usually hidden from the user-interface and ranked according to the semantic relevance between the query and the items.
In contrast, the recommended list reflects the final recommendations in the user interface, i.e., displaying on the right-hand panel of the video webpage.
On YouTube, the recommended list is a top-K sample from the concatenation of the relevant list, user demographics, watch history, search history, and spatial-temporal information \cite{covington2016deep}.
All features, apart from the relevant list, are user-, time- and location-dependent.
Hence, the displayed recommended list of the same video can be very different for two viewers, regardless of their logged-in state, location or viewing time.
On the other hand, the relevant list is consistent for all requests, from any client during any period of time.
We also observe the relevant list changes less frequently than the recommended list, which suggests it is more robust to the update of YouTube recommender systems.
For these reasons, we use the relevant list to construct and measure YouTube video network in \cref{sec:measures}.


\section{Measuring YouTube video network}
\label{sec:measures}

In this section, we present the macroscopic (\cref{ssec:macroscopic}), microscopic (\cref{ssec:microscopic}), and temporal (\cref{ssec:temporal}) profiling of the \vevo network.

\subsection{Macroscopic profiling of the \vevo network}
\label{ssec:macroscopic}

We first compute several basic statistics such as indegree distribution, view count distribution, and \vevo videos uploading trend (\cref{sssec:basic-statistics}).
Next, we study the connection between the network structure and video popularity (\cref{sssec:structure-viewcounts}).
Lastly, we use the bow-tie structure to characterize the \vevo network and we discuss the impact of different cutoff values (\cref{sssec:bowtie}).


\begin{figure}[tbp]
	\centering
	\subfloat{\includegraphics[width=1\textwidth]{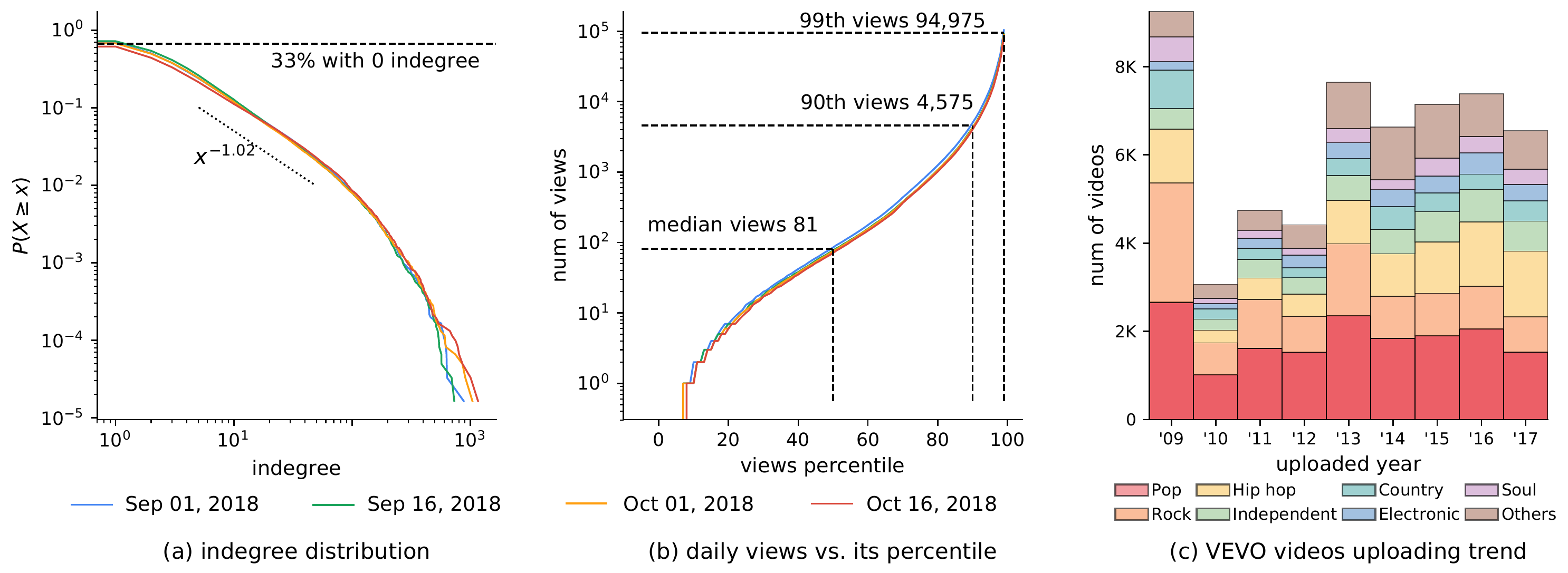}}
	\caption{
	\textbf{(a)} The indegree distribution of the \vevo network for four snapshots.
	The x-axis is the (log) indegree, and the y-axis is the (log) CCDF.
	On average, $33\%$ of all videos have no incoming links at the cutoff of 15.
	Best fitted power-law model is $x^{-1.02}$.
	\textbf{(b)} The average daily view counts distribution of \vevo network for the same four snapshots.
	The x-axis shows average daily view count percentiles, and the y-axis shows the raw number of view counts in log scale.
	Both \textbf{(a)} and \textbf{(b)} show that the macroscopic structure is temporally consistent.
	\textbf{(c)} The number of \vevo videos uploaded each year, broken down by genre. 
	}
	\label{fig:measure-statistics}
\end{figure}

\subsubsection{Basic statistics}
\label{sssec:basic-statistics}

\header{Over-represented medium-size indegree videos.}
Here we study the indegree distribution of the \vevo network.
Note that the outdegree of all nodes is bounded by the cutoff value on the relevant list and therefore not presented.
We remove all links pointing to non-\vevo videos, resulting in an average of 363,965 edges each day, and an average degree of 6.
Note that the average degree of 10 mentioned in \cref{sec:data} is obtained with a cutoff of 25, whereas here we study the relevant network constructed with a cutoff of 15, since the display probability of videos below position 15 appearing on recommended list is less than 0.32.
\cref{fig:measure-statistics}(a) shows the complementary cumulative density function (CCDF) of the indegree distribution for four different snapshots of the network, taken 15 days apart.
We notice that the indegree distribution does not resemble a straight line in the log-log plot, meaning it is not power-law, unlike for other online networks, e.g., the World Wide Web \cite{broder2000graph,meusel2014graph}, the network of interaction in online communities \cite{zhang2007expertise}, and the follower/following network on social media \cite{kwak2010twitter}.
The medium-sized indegree videos are over represented than that in the best fitted power-law model ($\alpha=2.02$, fitted by \textsc{powerlaw} package \cite{alstott2014powerlaw}, resulting in $x^{-1.02}$ in CCDF~\cite{clauset2009power}).
This result holds for all four snapshots.

\header{Attention is unequally allocated.}
\cref{fig:measure-statistics}(b) plots the average daily views against the view count percentile.
The daily view count at median is 81, but it is 4,575 at the 90th percentile.
These observations, together with a Gini coefficient of $0.946$, indicate that the attention allocation in the \vevo network is highly unequal --- the top 10\% most viewed videos occupy 93.1\% views.
We also find a moderate correlation between view count and indegree value (details in \cref{ssec:microscopic}).

\header{Uploading trend by music genre.}
To date, our dataset is the largest digital trace of \vevo artists on YouTube, allowing us to study the production dynamics of the \vevo platform.
\cref{fig:measure-statistics}(c) shows the number of \vevo videos that are uploaded each year from 2009 to 2017, broken down by their genres.
We omit year 2018 as we only observed 8 months for it (until August).
There is a significantly higher number of uploads (9,277) in 2009 as it is the year when \vevo was launched, and when many all-time favorite songs were syndicated to the YouTube platform.
Pop, Rock, and Hip hop music are the top 3 genres, accounting for 62.85\% of all uploads.
The \vevo videos upload rate is more or less constant around 7,000 since 2013.
The flattening production dynamics is somewhat surprising given the overall growth of YouTube~\cite{youtube2017billion}.

\subsubsection{Linking network structure and popularity}
\label{sssec:structure-viewcounts}


\begin{figure}[tbp]
	\centering
	\subfloat{\includegraphics[width=1\textwidth]{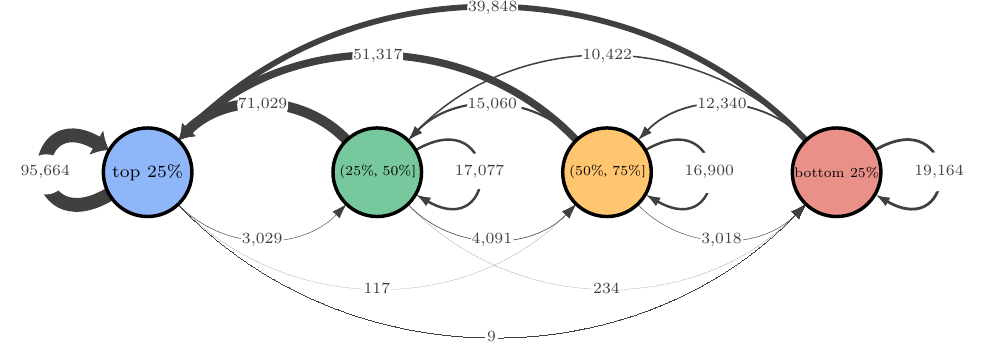}}
	\caption{
	The four video groups are constructed based on view count percentiles and the connections between them.
	All groups disproportionately point to more popular groups.
	}
	\label{fig:measure-connect}
\end{figure}

Here, we investigate the connection between the relevant network structure and video view counts.
Specifically, we divide the videos in the \vmg dataset into four equal groups by computing the view count quartiles.
Each group contains 15,185 videos.
Next, we count the number of edges that originate and end in each pair of groups.
\cref{fig:measure-connect} represents the four groups together with the number of links between them.
The ``top $25\%$'' group contains the top $25\%$ most viewed videos, while the ``bottom $25\%$'' contains the $25\%$ least viewed videos.
The width of the arrows is scaled by the number of the edges between the videos placed in the two groups.
One can conceptualize that the edges act as conduits for the attention to flow between different groups and their thickness indicate the probability that a random user jumps from one group to the other.
We observe that all four groups have the most links pointing to the ``top $25\%$'' group.
In fact, every group disproportionately points towards more popular groups than towards the less popular ones.
This means the recommendation network built by the platform is likely to take a random viewer towards more popular videos and keep them there, therefore reinforcing the ``rich get richer'' phenomenon.

\subsubsection{The bow-tie structure of video networks}
\label{sssec:bowtie}

The bow-tie structure was first proposed by~\citet{broder2000graph} to visualize the structure of the whole web.
It classifies the complex web graph into five components:
(a) the largest strongly connected component (LSCC) as the core; 
(b) the IN component which can reach the LSCC, but not the other way around; 
(c) the OUT component which can be reached from the LSCC, but not the other way around; 
(d) the Tendrils component which connect to either the IN or the OUT, bypassing the LSCC; 
(e) the Disconnected components which are disconnected from the rest of the components.
The strongly connected component (SCC) can be easily computed in linear time by using Tarjan's algorithm \cite{tarjan1972depth}.
For the \vevo network, we quantify the sizes of different components in the bow-tie structure using both the number of nodes (videos) and the amount of attention (views). 
Unlike the Web graph~\cite{broder2000graph}, we know the amount of views garnered by each video, this allows us to comparatively analyze the total attracted attention in each component.
The bow-tie structure is a good conceptual description, because the directed edges exist only from the IN to the LSCC component (similarly, LSCC to OUT, and IN to OUT) but not the other way around, indicating that the attention in the network can only flow in a single direction from IN to LSCC (similarly, LSCC to OUT, and IN to OUT).


\begin{table*}
  \caption{
  	Comparison of bow-tie structure in prior studies and \vevo network.
  }
  \label{tab:compare-bowtie}
  \resizebox{\textwidth}{!}{
  \begin{tabular}{r|rrrrrr}
    \toprule
    Component & Web '97 \cite{broder2000graph} & Web '12 \cite{meusel2014graph} & Forum~\cite{zhang2007expertise} & Citation~\cite{kim2012event} & \vevo videos & \vevo attention \\
    \midrule
    LSCC & 27.74\% & 51.28\% & 12.3\% & 4.26\% & 23.11\% & 82.60\% \\
    IN & 21.29\% & 31.96\% & 54.9\% & 54.93\% & 68.54\% & 12.74\% \\
    OUT & 21.21\% & 6.05\% & 13.0\% & 3.74\% & 0.35\% & 3.40\% \\
    Tendrils & 21.52\% & 4.87\% & 17.9\% & 24.76\% & 2.47\% & 1.13\% \\
    Disconnected & 8.24\% & 5.84\% & 1.9\% & 12.30\% & 5.53\% & 0.14\% \\
    \bottomrule
  \end{tabular}
  }
\end{table*}


\begin{figure*}[tbp]
	\centering
	\subfloat{\includegraphics[width=1\textwidth]{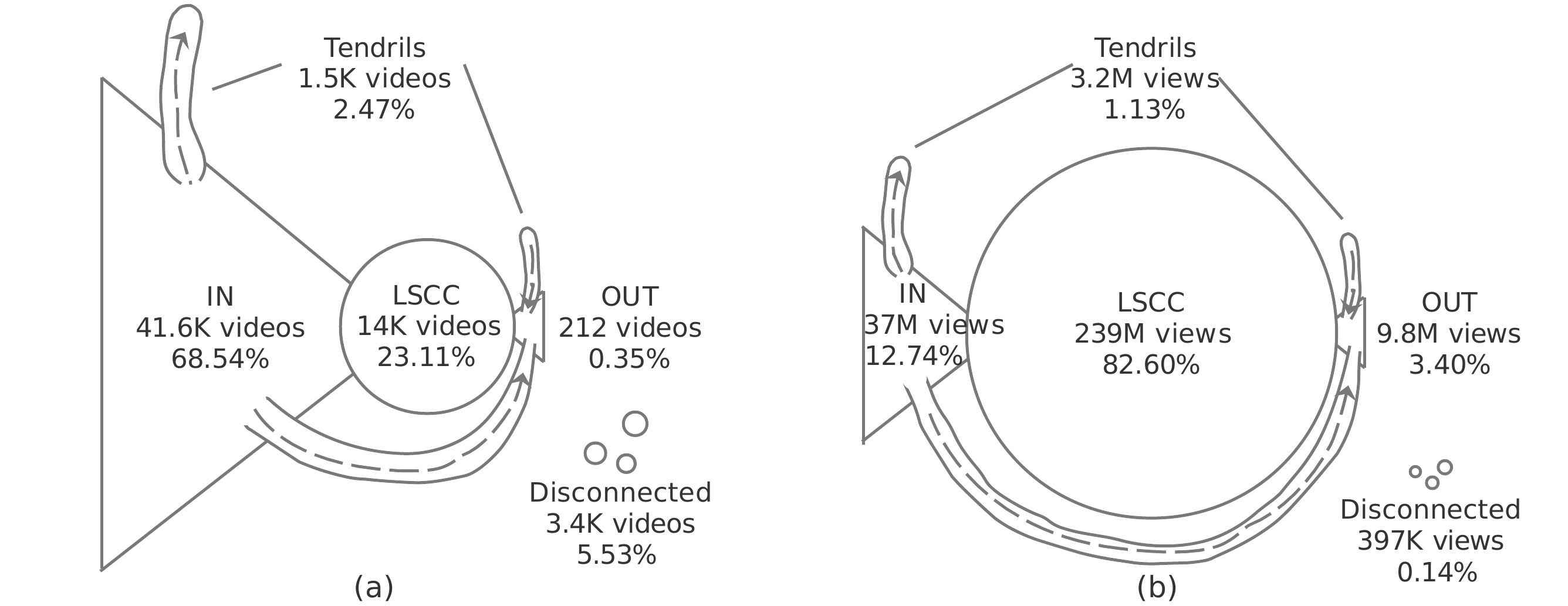}}
	\caption{
	\textbf{(a)} The bow-tie structure of the \vevo network, using a snapshot on Oct, 1, 2018 and a cutoff of 15 on the relevant list.
	\textbf{(b)} The bow-tie structure, with each component resized by its corresponding view counts. 
	The LSCC consumes the majority of attention in the \vevo network.
	}
	\label{fig:measure-bowtie}
\end{figure*}

\cref{tab:compare-bowtie} compares the relative sizes of each component in prior literature and in our \vevo network.
The \vevo network is quite different with respect to other previously studied online networks, e.g., the Web graph \cite{broder2000graph,meusel2014graph} and user activity network in online community~\cite{zhang2007expertise,kim2012event}.
It has a much larger IN component, encompassing 68.54\% of all the videos.
The OUT, Tendrils, and Disconnected components are all very small, accounting for a total of 8.35\% videos.
\cref{fig:measure-bowtie}(a) visualizes the bow-tie structure of the \vevo network.
Unlike other graphs, our \vevo graph is the by-product of the recommender systems, which is subjected to the proprietary algorithm and its updating cycle.
This suggests there may exist considerable temporal variation in the composition of the bow-tie components, see \cref{ssec:temporal} for observations over time.


\begin{figure*}[tbp]
	\centering
	\subfloat{\includegraphics[width=1\textwidth]{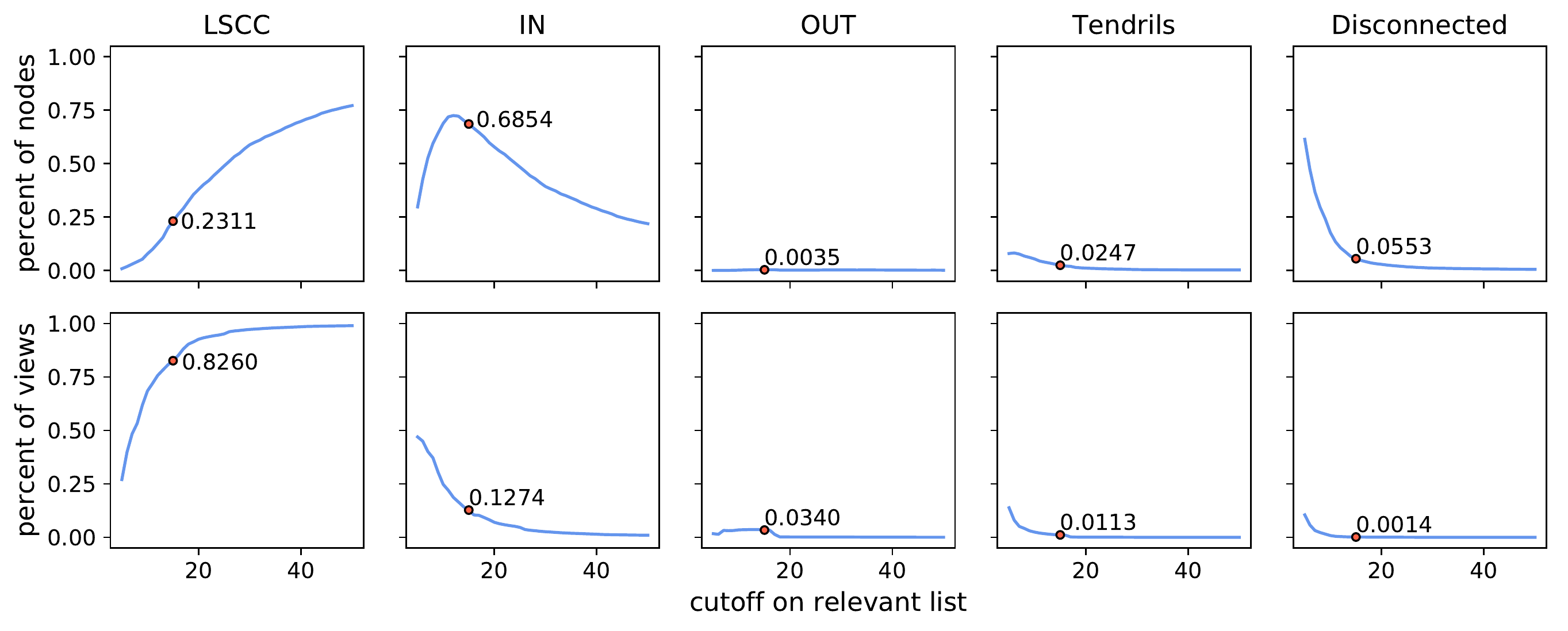}}
	\caption{
	The relative size of the components in bow-tie structure, as a function of the cutoff on the relevant list.
	Red dots denote the statistics at the cutoff of 15.
	}
	\label{fig:measure-bowtie-cutoff}
\end{figure*}

\cref{fig:measure-bowtie}(b) resizes each component of the \vevo bow-tie by the total view counts in it.
Visibly the roles of LSCC and IN are reversed: the LSCC now occupies $82.6\%$ attention (while accounting for only $23.11\%$ of the videos), while the big IN component ($68.54\%$ of the videos) only attract $12.74\%$ attention.
This is consistent with the observation in \cref{sssec:structure-viewcounts} that the attention is unequally allocated in the \vevo network.
Given the definition of the IN component, its $68.54\%$ of videos contribute attention towards the LSCC, but not the other way around (there is no link from LSCC towards IN).
As a result, the LSCC accumulates a large proportion of all attention.
The OUT, Tendrils, and Disconnected components account for almost negligible attention ($4.67\%$ of the views altogether).

\header{Impact of different cutoff values on the bow-tie structure.}
The \vevo network changes as we change the cutoff on the relevant list, as taking more edges into account densifies the network.
\cref{fig:measure-bowtie-cutoff} shows how the relative size of the bow-tie component changes with varying cutoff values.
As the cutoff increases, more edges are added to the network, especially for the videos in the Disconnected component.
Backwards links are formed between videos in the LSCC and IN, and as a result, the LSCC absorbs parts of the IN component.
Therefore, the LSCC increases, the IN decreases, while the other three components (OUT, Tendrils, and Disconnected) become negligible.
At cutoff of 50, the \vevo network structures into 2 distinct components: a LSCC component consisting of $77\%$ videos and $99\%$ attention, and an IN component consisting of the remaining $23\%$ videos and accounting for only $1\%$ of the attention.

\subsection{Microscopic profiling of the \vevo network}
\label{ssec:microscopic}

In this section, we jointly analyze the relation between video age, indegree, and popularity by examining overall correlation, as well as among top-ranked videos.


\begin{figure*}[tbp]
	\centering
	\subfloat{\includegraphics[width=0.96\textwidth]{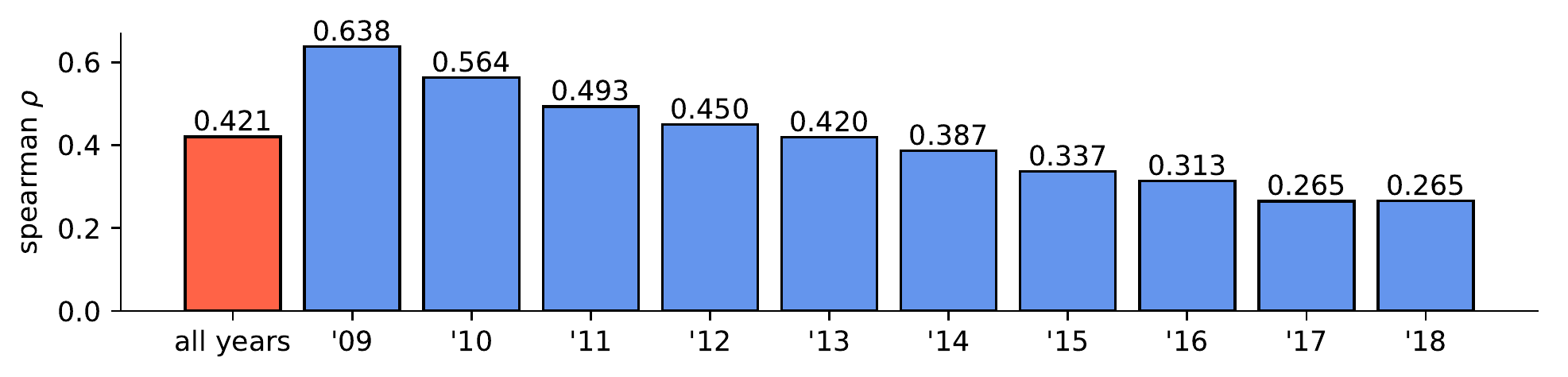}}
	\caption{Spearman's $\rho$ between video indegree and views, across all videos or disaggregated by uploaded year.}
	\label{fig:measure-spearmanr}
\end{figure*}

\header{The disconnect between network indegree and video view count.}
We measure the correlation between video indegree and view count using Spearman's $\rho$ --- a measure of the strength of the association between two ranked variables, and which takes values between -1 and +1.
A positive $\rho$ implies the ranks of the two variables move together in the same direction.
At the level of the entire dataset, we detect a moderate correlation between video indegree and view count (Spearman's $\rho = 0.421^{***}$, p < 0.001).
\cref{fig:measure-spearmanr} shows the Spearman's $\rho$ when we further break down the videos in the \vmg based on their uploaded year.
We observe that the strength of the correlation decreases for fresher videos.
Videos uploaded in 2009 have a much stronger correlation ($\rho = 0.638^{***}$) than videos uploaded in 2018 ($\rho = 0.265^{***}$).
This suggests that video age is an important confounding factor when one tries to estimate the effects of the recommendation network.
Empirically, this may indicate the shift in what drives attention towards video consumption.
\citet{zhou2010impact} have measured that the two main drivers for video views are YouTube search and recommender.
One explanation of our observation above is that as videos get older, the effects of recommendation become more pronounced.

\header{A closer look at the top videos.}
\cref{tab:top20} presents the top 20 videos with highest average daily indegree (top panel) and top 20 videos with highest average daily views (bottom panel).
We observe a modest amount of discrepancy between these two dimensions, with only 5 videos being on both lists (shown in bold font).
Most of the top-viewed videos are relatively new to the platform --- 10 out 20 are published within one year and the top 5 are all within the past 7 months (relative to November 2018).
In contrast, the videos with high indegree are mostly songs with sustained interests, some dating back to 10 years ago, such as ``The Cranberries - Zombie'' and ``Bon Jovi - It's My Life''.
These two songs were respectively released in 1993 and 2000, having existed for a long time before being uploaded to YouTube.
Currently, they still attract half a million views everyday after nearly 20 years, ranking 3rd and 17th on the most-linked video list, respectively.
This may shed light onto why video popularity lifecycle exhibits a multi-phase pattern~\cite{yu2015lifecyle}.
Our observations do not conflict with the design of YouTube recommender systems, which promote ``reasonably recent and fresh'' content \cite{davidson2010youtube,covington2016deep,beutel2018latent}.
Fresh videos can be recommended due to the relevance, novelty and diversity trade-offs~\cite{konstan2012recommender,ziegler2005improving}.
Instead, our observed video relations are based on the content recommendation network~\cite{carmi2017oprah,dhar2014prediction}.


\begin{table*}
  \caption{
	Top 20 most-linked (top panel) and top 20 most-viewed videos (bottom panel). 
  	Both the indegree and the view counts are the average of daily values across 9 weeks. 
  	The age (in days) is calculated till Nov 2, 2018. 
  	Only 5 videos appear in both charts (\textbf{boldfaced}).
  	Most high indegree videos are songs with sustained interests, whereas most highly viewed videos are recently uploaded.
  }
  \label{tab:top20}
  \resizebox{\textwidth}{!}{
  \begin{tabular}{rrrrrrr}
    \toprule
    Video title & Artist & Age & Indegree & -rank & Views & -rank \\
    \midrule
    \textbf{Girls Like You} & \textbf{Maroon 5} & \textbf{155} & \textbf{870} & \textbf{1} & \textbf{7,167,077} & \textbf{1} \\
    Rolling in the Deep & Adele & 2,894 & 835 & 2 & 703,495 & 42 \\
    Zombie & The Cranberries & 3,426 & 769 & 3 & 580,928 & 66 \\
    \textbf{Something Just Like This} & \textbf{Chainsmokers} & \textbf{618} & \textbf{732} & \textbf{4} & \textbf{1,840,077} & \textbf{6} \\
    \textbf{Counting Stars} & \textbf{OneRepublic} & \textbf{1,981} & \textbf{714} & \textbf{5} & \textbf{1,632,001} & \textbf{9} \\
    \textbf{Uptown Funks} & \textbf{Mark Ronson} & \textbf{1,444} & \textbf{587} & \textbf{6} & \textbf{1,724,938} & \textbf{7} \\
    Here Without You & 3 Doors Down & 3,315 & 541 & 7 & 401,989 & 114 \\
    Someone Like You & Adele & 2,591 & 514 & 8 & 954,981 & 26 \\
    Mr. Brightside & The Killers & 3,426 & 509 & 9 & 197,313 & 255 \\
    The Pretender & Foo Fighters & 3,317 & 480 & 10 & 266,610 & 182 \\
    I Want It That Way & Backstreet Boys & 3,296 & 464 & 11 & 368,859 & 131 \\
    Unforgettable & French Montana & 587 & 463 & 12 & 682,396 & 44 \\
    \textbf{Dusk Till Dawn} & \textbf{ZAYN} & \textbf{421} & \textbf{452} & \textbf{13} & \textbf{1,011,255} & \textbf{20} \\
    Starboy & The Weeknd & 765 & 450 & 14 & 519,343 & 77 \\
    Hello & Adele & 1,106 & 447 & 15 & 518,429 & 78 \\
    Love The Way You Lie & Eminem & 3,011 & 433 & 16 & 729,344 & 37 \\
    It's My Life & Bon Jovi & 3,425 & 426 & 17 & 470,175 & 91 \\
    Cups & Anna Kendrick & 2,030 & 419 & 18 & 140,614 & 386 \\
    Say You Won't Let Go & James Arthur & 784 & 417 & 19 & 774,130 & 32 \\
    Pumped up Kicks & Foster The People & 2,827 & 408 & 20 & 420,350 & 107 \\
    \midrule
    \midrule
    \textbf{Girls Like You} & \textbf{Maroon 5} & \textbf{155} & \textbf{870} & \textbf{1} & \textbf{7,167,077} & \textbf{1} \\
    Sin Pijama & Becky G & 196 & 28 & 2,411 & 3,988,681 & 2 \\
    Taste & Tyga & 170 & 242 & 81 & 2,542,673 & 3 \\
    Cuando Te Bese & Becky G & 92 & 0 & 40,040 & 2,373,613 & 4 \\
    Rise & Jonas Blue & 140 & 40 & 1,603 & 1,937,467 & 5 \\
    \textbf{Something Just Like This} & \textbf{Chainsmokers} & \textbf{618} & \textbf{732} & \textbf{4} & \textbf{1,840,077} & \textbf{6} \\
    \textbf{Uptown Funks} & \textbf{Mark Ronson} & \textbf{1,444} & \textbf{587} & \textbf{6} & \textbf{1,724,938} & \textbf{7} \\
    No Tears Left to Cry & Ariana Grande & 196 & 84 & 597 & 1,634,916 & 8 \\
    \textbf{Counting Stars} & \textbf{OneRepublic} & \textbf{1,981} & \textbf{714} & \textbf{5} & \textbf{1,632,001} & \textbf{9} \\
    Thunder & Imagine Dragons & 548 & 72 & 764 & 1,474,712 & 10 \\
    One Kiss & Calvin Harris & 184 & 23 & 2,973 & 1,230,173 & 11 \\
    Natural & Imagine Dragons & 70 & 8 & 7,747 & 1,186,965 & 12 \\
    Believer & Imagine Dragons & 605 & 60 & 998 & 1,174,431 & 13 \\
    Mayores & Becky G & 476 & 62 & 958 & 1,173,191 & 14 \\
    What's Up & 4 Non Blondes & 2,809 & 355 & 36 & 1,128,159 & 15 \\
    Sugar & Maroon 5 & 1,388 & 349 & 37 & 1,116,300 & 16 \\
    God's Plan & Drake & 258 & 227 & 102 & 1,093,048 & 17 \\
    Sicko Mode & Travis Scott & 91 & 64 & 913 & 1,076,694 & 18 \\
    Whatever It Takes & Imagine Dragons & 386 & 118 & 347 & 1,016,394 & 19 \\
    \textbf{Dusk Till Dawn} & \textbf{ZAYN} & \textbf{421} & \textbf{452} & \textbf{13} & \textbf{1,011,255} & \textbf{20} \\
    \bottomrule
  \end{tabular}
  }
\end{table*}

Another group of interest is the videos that are highly viewed yet with low indegree.
We find this pattern appears at the level of the artist.
For instance, ``Becky G'' has 3 videos on the top 20 most-viewed list, ranking 2, 4, and 14.
However, the indegrees for her videos are extremely low (rank 2411, 40040, and 958 respectively).
Particularly, the video ``Cuando Te Bese'' attracts an average of 2.4M views every day for 9 consecutive weeks.
However, it has only one video pointing to it from the rest of the 60,739 \vevo videos.
A closer look reveals that ``Becky G'' is an American singer who often releases Spanish songs.
The above observation shows that her videos are either recommended from non-English and/or non-\vevo videos, e.g., the Spanish songs community, or that recommendation network is not the main traffic driver for her videos.

\subsection{Temporal evolution of \vevo network}
\label{ssec:temporal}

Here, we study the dynamics of the \vevo network over 9 weeks, namely the appearance and disappearance of recommendation links between videos.
We show that pairs of videos can have either ephemeral link or frequent link between them.


\begin{figure}[tbp]
	\centering
	\subfloat{\includegraphics[width=1\textwidth]{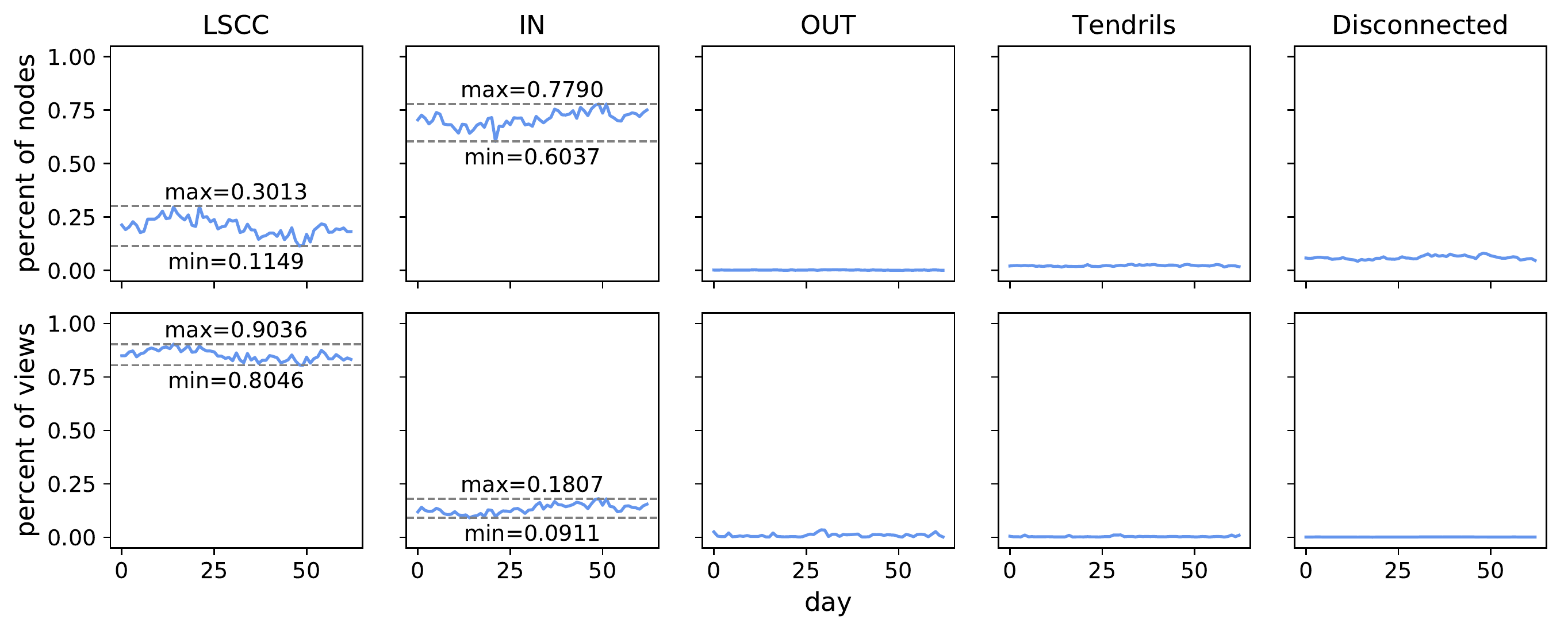}}
	\caption{Temporal evolution of the bow-tie structure over 63 days.}
	\label{fig:measure-temporal-macro}
\end{figure}

\header{Macroscopic dynamics.}
\cref{fig:measure-statistics}(a) and (b) show that both the indegree distribution and the view count distribution are temporally consistent.
However, when we plot the size variation of the different components in the bow-tie structure, we obtain a more nuanced story.
\cref{fig:measure-temporal-macro} shows that the size of the LSCC ranges from 11.49\% to 30.13\%, while IN component from 60.37\% to 77.9\% over 9 weeks.
Similarly, the percentage of total views in the LSCC ranges from 80.46\% to 90.36\%, while IN component from 9.11\% to 18.07\%.
Given that the same set of videos is tracked throughout the observation period and no new video is added, the above observations imply a significant turnover in the recommendation links between videos.
For example, the appearance of a link will allow a node to transition from the IN to the LSCC component; the disappearance of the same link would make it drop back into IN component.

\begin{figure}[tbp]
	\centering
	\subfloat{\includegraphics[width=1\textwidth]{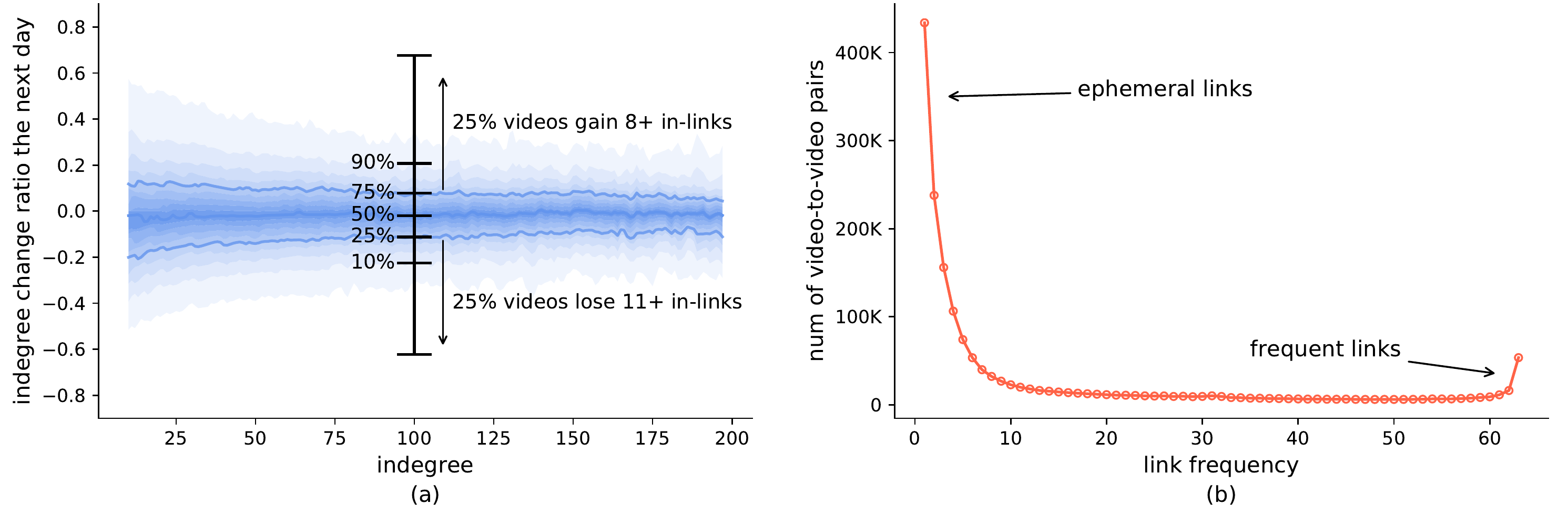}}
	\caption{
		\textbf{(a}) Daily indegree change rate for videos that have at least 20 in-links.
		\textbf{(b)} Link frequency of video-to-video pairs. 
		434K (25.2\%) links appear only once while 54K (3.1\%) links appear every day of our 9-week observation windows.}
	\label{fig:measure-temporal-micro}
\end{figure}

\header{Incoming ego-network dynamics.}
We study the link turnover using the incoming ego-network for each video.
Ego network consists of an individual focal node and the edges pointed towards it.
We only consider incoming edges, as the number of outgoing edges is capped by the relevant list cutoff (here the cutoff is 15).
For each video, we first extract the days with at least 20 incoming links.
Then for each day $t$, we compute the indegree change ratio between day $t$ and day $t+1$ by dividing the indegree delta (positive or negative) by the value in day $t$.
We obtain a number between -1 and 1, where -1 means that the video loses all of its incoming edges, and a value of 1 signifies that the video doubles the number of incoming edges.
\cref{fig:measure-temporal-micro}(a) shows the indegree change ratio summarized as quantiles, broken down by the value of indegree.
We highlight the 10th, 25th, median, 75th, and 90th percentile for the videos with an indegree of $100$.
$25\%$ videos with an indegree of $100$ will gain at least 8 in-links on the next day while another 25\% lose at least 11 in-links.
The median is around zero, meaning that there are as many videos that gain links as these that lose links.
Overall, this suggests that videos have very dynamic incoming ego-networks, with a non-trivial number of edges prone to appear and disappear from one day to another.


\header{Ephemeral links and frequent links.}
Given the rate at which links appear and disappear, here we ask the question if there exist videos that are frequently connected.
For each pair of connected videos, we count the number of times that a link appears between them over the 63 daily snapshots.
\cref{fig:measure-temporal-micro}(b) plots the link frequency (taking values between 1 and 63) on the x-axis and the number of video-to-video pairs with that link frequency on the y-axis.
We find that many links are ephemeral --- they appear several times, scattering in the 63 days time window.
We count that 434K ($25.2\%$) video-to-video links only appear once.
On the other hand, there are links that appear in every snapshot --- we count 54K ($3.1\%$) such links.
Ephemeral links may contribute to bursty popularity dynamics of YouTube videos, and to the generally perceived unpredictability in complex social systems~\cite{martin2016exploring,rizoiu2017expecting,rizoiu2018sir}.
Frequent links may hold the answer to understanding and predicting the attention flow in a network of content.


\section{Estimating attention flow in the YouTube video network}
\label{sec:models}

The goal of this section is to estimate how well can the view counts of a video $v$ at day $t$ (denoted by $\mathbf{y}_v[t]$) be predicted, given (1) the view series of $v$ in the past $w$ days, $\mathbf{y}_{v}[t-w],\ldots\mathbf{y}_{v}[t-1]$; (2) the view series, $\mathbf{y}_{u}[t-w],\ldots\mathbf{y}_{u}[t]$, for the set of videos $\{u | (u \rightarrow v) \in G \}$ pointing to $v$.

To this end, we first define and extract a persistent network that contains links appearing throughout all the snapshots (\cref{ssec:persistent}).
Next, we detail the setup of predicting video popularity with recommendation network information (\cref{ssec:setting}).
We analyze the prediction results and provide an analysis on the strength of each link (\cref{subsec:prediction-results}).
Finally, we introduce a new metric --- estimated network contribution ratio.
We use it to identify the types of content that benefit most from being recommended in the network (\cref{subsec:result-interpretation}). 

\subsection{Constructing a network with persistent links}
\label{ssec:persistent}

In order to reliably estimate the effects of the recommendation network on the viewing behaviors, we apply two filters: 
(a) target videos should have at least 100 daily views on average;
(b) the average daily views of the source videos should be at least 1\% of those of the target videos as such videos cannot substantially influence their far more popular neighbors.
In the resulting network, we further remove the \textit{ephemeral links} that appear sporadically over time and correct for the \textit{missing links} that appear frequently, but with scattered gaps in between their appearances.
We assume that the missing links are likely to exist in the scattered gaps, and we use a majority smoothing method to find them (detailed next).
Links appearing in all the 63 daily snapshots and the corrected missing links, both dubbed \textit{persistent links}, make up the \textit{persistent network}.

\header{Finding persistent links.}
We use a moving window of length 7, same as the weekly seasonality, to extract the persistent structure of the \vevo network over the 63-day observation window.
A link from video $u$ to video $v$, $(u \rightarrow v)$, is maintained on day $t$ if $(u \rightarrow v)$ appears in a majority ($\geq 4$) of the days in time window $[t-3, t+3]$.
Likewise, if a link is missing on the current day $t$ but it appears in the majority of surrounding 7-day window, we consider it is a missing link and add it back to the network.
When $t-3$ is earlier than the first day of data collection, or $t+3$ later than the last day, we still apply the majority rule on the available days.
The resulting graph has 52,758 directed links, pointing from 28,657 source videos to 13,710 target videos.
Among them, 2,696 links are reciprocal, meaning two videos mutually recommend each other.
We find significant homophily in the persistent network:
33,908 (64.3\%) links have both the source and the target videos belonging to the same \vevo artist, and 44,154 (83.7\%) links are between videos of the same music genre.


\begin{figure*}[tbp]
	\centering
	\subfloat{\includegraphics[width=1\textwidth]{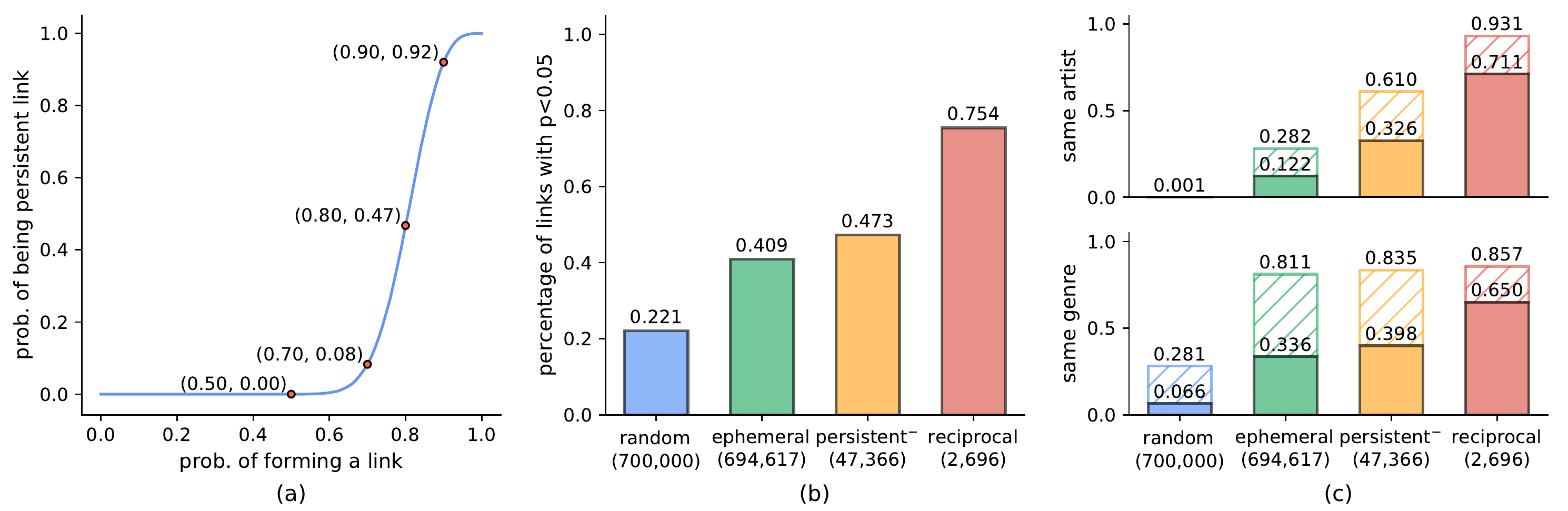}}
	\caption{
	\textbf{(a)} The probability of forming a persistent link (y-axis) as a function of the probability of forming a link (x-axis). 
	\textbf{(b)} Fraction of statistically correlated links in four groups at significance level of 0.05. 
	The numbers in the brackets indicate the number of links in each group.
	\textbf{(c)} The numbers above the shaded bar indicate the fraction of links between videos from the same artist (top) or with the same genre (bottom). 
	The numbers above the solid bar indicate the fraction of links connecting videos with the same artist/genre and whose popularity dynamics are statistically correlated at significance level of 0.05.
	}
	\label{fig:model-persistent}
\end{figure*}

\header{Validating persistent links via simulation.}
We illustrate the probability of persistent links by simulating a simple link presence/absence model.
We assume a link is independently presented on each day with probability $p_l \in [0, 1]$, and absent with probability $1 - p_l$.
We first simulate the link formulation for 63 times, then apply our 7-day majority smoothing to determine if it is persistent.
We repeat the simulation for 100,000 times, and compute the probability of a link being persistent, denoted by $\xi$.
In \cref{fig:model-persistent}(a), we plot the obtained $\xi$ against varying $p_l$.
For $p_l=0.5$ the edge is never persistent ($\xi = 0$), whereas for $p_l=0.9$ the edge is very likely to be persistent ($\xi = 0.92$).
From the simulation results, we can see that our 7-day majority smoothing rule favors links that appear much more frequent than chance, and suppresses links that appear lower or closer to chance.

\header{Videos connected by persistent links have correlated popularity dynamics.}
We use Pearson's $r$ to measure the correlation between the popularity dynamics of two videos connected by a persistent link.
It is known that the cross-correlation of time series data is affected by the within-series dependence.
Therefore, we deseasonalize, detrend, and normalize the view count series by following the benchmark steps in the M4 forecasting competition~\cite{m4forecasting}.
This is to ensure that the residual time series data is stationary and to avoid spurious correlations.
We compute the Pearson's $r$ on the obtained residual data, and we perform a paired correlation test which we consider statistically significant for $p < 0.05$.

\cref{fig:model-persistent}(b) shows the fraction of links for which the correlation test is statistically significant over four groups of links.
The $persistent^{-}$ group contains all the 52,758 persistent links we identified but excluding the 2,696 pairs of \textit{reciprocal} links --- resulting in 47,366 persistent yet non-reciprocal links.
The \textit{ephemeral} group consists of all links which have been deemed as non-persistent after applying the 7-day majority smoothing.
The \textit{random} group is constructed by randomly selecting pairs of unconnected videos and pretending that they have a link.
All groups are filtered based on the same two criteria mentioned before.
There are a total of 694,617 links in the ephemeral group and we sample 700,000 links in the random group.
We find that $75.4\%$ of the reciprocal links connect videos with statistically correlated popularity series.
We include both positive and negative correlations as two user attention series may cooperate or compete with each other~\cite{zarezade2017correlated}.
Combining the reciprocal and persistent$^{-}$ groups, 26,460 (50.2\%) links in our persistent network have correlated dynamics.
This is much higher than the percentage for ephemeral links (40.9\%) and that for unconnected random video pairs (22.1\%).

We further examine the content similarity in the persistent links by grouping links that connect videos from the same artist or with the same music genre (described in~\cref{fig:measure-statistics}(c)).
\cref{fig:model-persistent}(c) top shows that most reciprocal links (93.1\%) connect videos from the same artist, while 71.1\% of them have statistically correlated popularity dynamics.
The percentages are slightly lower for the persistent$^{-}$ group (61\% from the same artist, and 32.6\% with correlated popularity) and it drops even lower for ephemeral group (28.2\% and 12.2\%, respectively).
The situation is slightly different when we study the links that connect videos of the same genre, as shown in \cref{fig:model-persistent}(c) bottom.
We find that more than 80\% of the links connect videos of the same genre, irrespective of whether they are sporadically or persistently connected.
The percentages of statistically correlated links with the same genre follow the same trend as those from the same artist, i.e., highest for reciprocal (65\%), followed by persistent$^{-}$ (39.8\%), ephemeral (33.6\%) and lowest for random (6.6\%).
The above observations indicate that not all persistent links have the same effect on video popularity, and motivate us to build a prediction model for each of the links.

\subsection{Prediction setup and models}
\label{ssec:setting}

\header{Prediction setting.}
One important observation is that viewing dynamics exhibit a 7-day seasonality~\cite{huang2018user,cheng2008statistics}.
In our temporal hold-out setting, we use the first 8 weeks (2018-09-01 to 2018-10-26) to train the model and we predict the daily view counts in the last week (2018-10-27 to 2018-11-02).
This chronological split ensures that the training data temporally precedes the testing data.
If at any point we are required to use the day $t+1$ to predict the day $t+2$ (when both $t+1$ and $t+2$ are in the testing period), we use the predicted value $\hat{\mathbf y}[t+1]$ instead of observed value $\mathbf{y}[t+1]$.

\header{Evaluation metric.}
The predicting performance is quantified using the symmetric mean absolute percentage error (SMAPE).
SMAPE is an alternative to the mean absolute percentage error (MAPE) that can handle the case when the true value or the predicted value is zero.
It is a scale-independent metric and suitable for our task in which the volume of views for different videos vary considerably.
Formally, SMAPE can be defined as

\begin{equation}
  \label{eq:smape}
  \mathrm{SMAPE}(v) = \frac{200}{\mathbf T} \sum_{t=1}^{\mathbf T}\frac{|\mathbf{y}_v[t] - \hat{\mathbf{y}}_v[t]|}{|\mathbf{y}_v[t]| + |\hat{\mathbf{y}}_v[t]|}
  \quad\text{or}\quad
  \mathrm{SMAPE}(t) = \frac{200}{|G|} \sum_{v \in G}\frac{|\mathbf{y}_v[t] - \hat{\mathbf{y}}_v[t]|}{|\mathbf{y}_v[t]| + |\hat{\mathbf{y}}_v[t]|}
\end{equation}

where $\mathbf{y}_v[t]$ is the true value for video $v$ on day $t$, $\hat{\mathbf{y}}_v[t]$ is the predicted value, $\mathrm{T}$ is maximal forecast horizon, and $G$ is the persistent network.
SMAPE$(v)$ averages the forecast errors over different horizons for an individual video $v$, while SMAPE$(t)$ averages over different videos for a certain forecast horizon $t$.
The overall SMAPE for each model is computed by taking the arithmetic mean of SMAPEs over different horizons and over all videos.
SMAPE ranges from 0 to 200, while 0 indicates perfect prediction and 200 the largest error, when one of the true or the predicted values is 0.
When the true and the predicted are both 0, we define SMAPE to be 0.

\header{Baseline models.}
We use a few off-the-shelf time series forecasting methods from naive forecast to recurrent neural network.
The baseline models are estimated on a per-video basis.

\begin{itemize}[leftmargin=*]
  \item {Naive}: The forecast at all future times is the last known observation.

\begin{equation}
  \label{eq:naive}
    \hat{\mathbf{y}}_v[t] = \mathbf{y}_v[\mathrm{T}^*]
\end{equation}
  where $\mathrm{T}^*$ is the last day in the training phase.

  \item {Seasonal naive (SN)}: The forecast is the corresponding observation in the last seasonal cycle.
  This method often works well for seasonal data.
  We observe that many videos in the \vmg dataset exhibit a 7-day seasonality.
  Therefore we set the periodicity length $\mathrm{m}^*$ to be 7.

\begin{equation}
  \label{eq:snaive}
    \hat{\mathbf{y}}_v[t] = \mathbf{y}_v[t - \mathrm{m}^*]
\end{equation}
  
  \item {Autogressive (AR)}: AR is one of the most commonly used model in time series forecasting.
  An AR model of order $p$ describes the relation between each of the past $p$ days and current day, formally defined as:

\begin{equation}
  \label{eq:ar}
    \hat{\mathbf{y}}_v[t] = \sum_{\tau = 1}^{p} \alpha_{v, \tau} {\mathbf{y}_v}[t-\tau]
\end{equation}
  We choose the order $p$ to be 7.
  $\alpha_{v, \tau}$ represents the relation between current day and $\tau$ days before.
	
  \item {Recurrent neural network (RNN)}: RNN is a deep learning architecture that models temporal sequences.
  We implement RNN with long short-term memory (LSTM) units.
  LSTM-based approaches have been competitive in time series forecast tasks, mainly in a sequence-to-sequence (seq2seq) setup, see \cite{kuznetsov2019foundations} for detailed discussions.
\end{itemize}

\header{Networked popularity model.}
Built on top of the AR model, we model the network effects by assigning a weight $\beta_{u, v}$ to each link $(u \rightarrow v)$ existing in the persistent graph $G$, which modulates the inbound traffic received via that link, defined as:

\begin{equation}
  \label{eq:network}
    \hat{\mathbf{y}}_{v}[t] = \sum_{\tau = 1}^{p} \alpha_{v, \tau} {\mathbf{y}_{v}}[t-\tau] + \sum_{(u, v) \in G} \beta_{u, v} \mathbf{y}_{u}[t]
\end{equation}

$\beta_{u, v}$ can be explained as the probability that a generic user clicks on video $v$ from video $u$, therefore, we impose the constraint $0 \leq \beta_{u, v} \leq 1$.
We refer to this model as ARNet.

One way to interpret the ARNet is to conceptualize a YouTube watching session as a sequence of video clicking.
We therefore categorize views on YouTube into two classes: \textit{initial} views and \textit{subsequent} views.
The initial views start the clicking sequences.
Some possible entry points include homepage feed, search results, or YouTube URLs on other social media.
The subsequent views model the behaviors of users clicking by following the recommendation links.
The session ends when the user navigates back to YouTube homepage, or quits the browser.
Although in the dataset we cannot differentiate initial views from subsequent views, we consider that initial views are driven by the latent interest of users, modelled as autoregression of the past $p$ days; in contrast, subsequent views are directed by the recommendation network, modelled as contribution from its incoming neighbours $\{u | (u \rightarrow v) \in G \}$ and mediated by estimated link strength $\beta_{u, v}$.

We use the \textsc{statsmodels.tsa} package for the AR model, \textsc{keras} package for the RNN, and build a customized optimization task with constrained \textsc{L-BFGS} for the ARNet.
We use the SMAPE as objective function in both RNN and ARNet.

\subsection{Popularity prediction results}
\label{subsec:prediction-results}

\begin{figure*}[tbp]
	\centering
 	\subfloat{\includegraphics[width=1\textwidth]{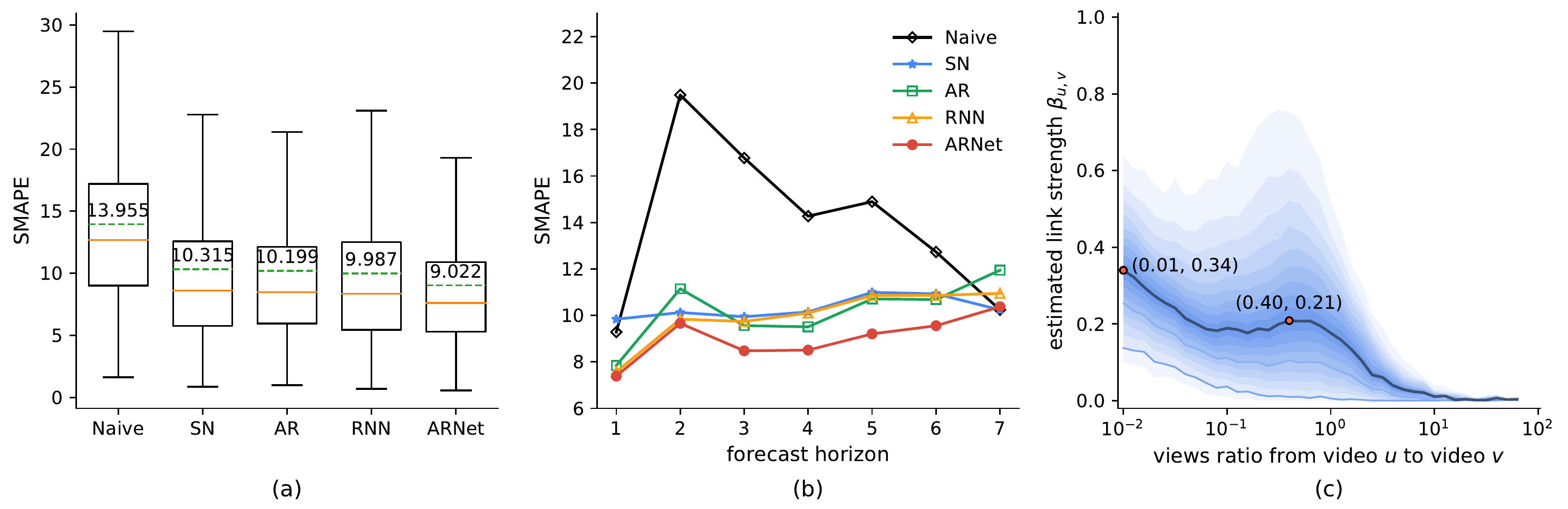}}
	\caption{Summary of prediction results, SMAPE: lower is better. 
	    \textbf{(a)} Boxplots aggregate the prediction performances over the 13,710 videos in the test set.
	    The dotted green line and the values show the mean SMAPE.
		\textbf{(b)} SMAPE for different forecast horizons (in days). 
		\textbf{(c)} The distribution of estimated link strength $\beta_{u, v}$ (y-axis) against the ratio of views of source video to that of target video (x-axis, in log scale). It has a bi-modal shape. 
	}
	\label{fig:model-results}
\end{figure*}

\cref{fig:model-results}(a) summarizes the prediction errors achieved by the five methods defined in \cref{ssec:setting}.
The Naive model alone is a weak predictor, however accounting for the seasonal effects (SN model) yields a significant error decrease.
It is worth noticing that the AR model yields similar performance as the advanced RNN model --- due to the known result that future popularity of online videos correlates with their past popularity \cite{pinto2013using}.
We observe that using recommendation network information further improves the prediction performance:
the ARNet model achieves a 9.66\% relative error reduction compared to the RNN model.
This prediction task shows that one can better predict the view series for a video if the list of videos pointing to it is known.
Next we study the prediction performance with respect to the forecast horizon, i.e., how many days in advance do we predict.
We average the SMAPEs over all videos against predictions for a given forecast horizon $t$, computed as $\mathrm{SMAPE}(t)$ in \cref{eq:smape}.
\cref{fig:model-results}(b) shows a nuanced story: the prediction performances decrease for all models as the forecast horizon extends.
Nevertheless, the ARNet model consistently outperforms other baselines across all forecast horizons, especially for larger horizons.


We posit two factors in preventing the models from obtaining even better results.
Firstly, it is well known that the attention dynamics tend to be bursty when items are first uploaded~\cite{rizoiu2017online,cheng2016cascades,martin2016exploring}, and the interest dissipates with time~\cite{figueiredo2016trendlearner}.
Given that 56,845 (93.6\%) videos in our dataset have been uploaded for more than one year and 9,277 (15.3\%) videos for almost ten years, most of the videos have passed the phases of the initial attention burst.
As a result, a large part of popularity variation comes from the weekly seasonality, rendering the simple seasonal naive model particularly competitive when compared to the more advanced RNN method.
The second is data sparsity when we build the models on a per-video basis.
RNN works best when it has ample volumes of data to train.
However, we use a sliding 7-day windows to predict the views in the next 7 days as suggested in~\cite{kuznetsov2019foundations}, therefore our data size is limiting the effectively training of the RNN model.

In our ARNet model, the estimated link strength $\beta_{u, v}$ can be used to quantify the influence from a video to its neighbours.
In \cref{fig:model-results}(c), we plot the distribution of $\beta_{u, v}$ against the ratio of views of source video to that of target video.
We split the x-axis into 40 equally wide bins in log scale.
Within each bin, we compute the values at each percentile, and then connect the same percentile across all bins.
The median line is highlighted in black.
The lighter the color shades are, the further the corresponding percentiles are away from the median.
We observe the distribution has a bi-modal shape with the first mode in 0.01 and second in 0.40 (for the median), meaning users are more likely to click a much more popular video (100 times more popular), or a moderate more popular video (2.5 times).
In contrast, the estimated link strength towards a less popular video is very low.
This observation, together with the measurement that videos disproportionately point to more popular videos (\cref{sssec:structure-viewcounts}), further reinforces the ``rich get richer'' phenomenon.

\subsection{The impacts of network on video popularity prediction}
\label{subsec:result-interpretation}

From the ARNet model, we derive a metric called the estimated network contribution ratio $\eta_v$, which is defined as

\begin{equation}
  \label{eq:net-ratio}
  \eta_v = \frac{\sum_{t=1}^{\mathbf T} \sum_{(u, v) \in G} \beta_{u, v} \mathbf{y}_{u}[t]}{\sum_{t=1}^{\mathbf T} \hat{\mathbf{y}}_{v}[t]}
\end{equation}
$\eta_v$ is the fraction of estimated inbound traffic from video $v$'s neighbours against its own predicted popularity.
As we constrain all coefficients in \cref{eq:network} to be non-negative, $\eta_v$ is bounded in $[0, 1]$. 
In our dataset, the mean $\eta_v$ is $0.314$.
In other words, for an average video in the \vmg dataset, 31.4\% of its views are estimated from the recommendation network.
This value is slightly higher than the YouTube network contribution measured by \citet{zhou2010impact} in 2010 (reported below 30\%).
We posit two potential reasons: (1) the \vevo network is more tightly connected than a random YouTube video network~\cite{airoldi2016follow}; (2) traffic on recommendation links may have increased since then, signifying the advances of modern recommender systems.
Furthermore, among the 31.4\% networked views, 85.9\% are estimated from the same artist, echoing the network homogeneity found by \citet{airoldi2016follow}.
On average, the 13,710 target videos in the persistent network attract 245.3M views every day.
Our ARNet model estimates that 78.6M (32\%) of these views are contributed via the recommendation network.


\begin{figure*}[tbp]
	\centering
 	\subfloat{\includegraphics[width=1\textwidth]{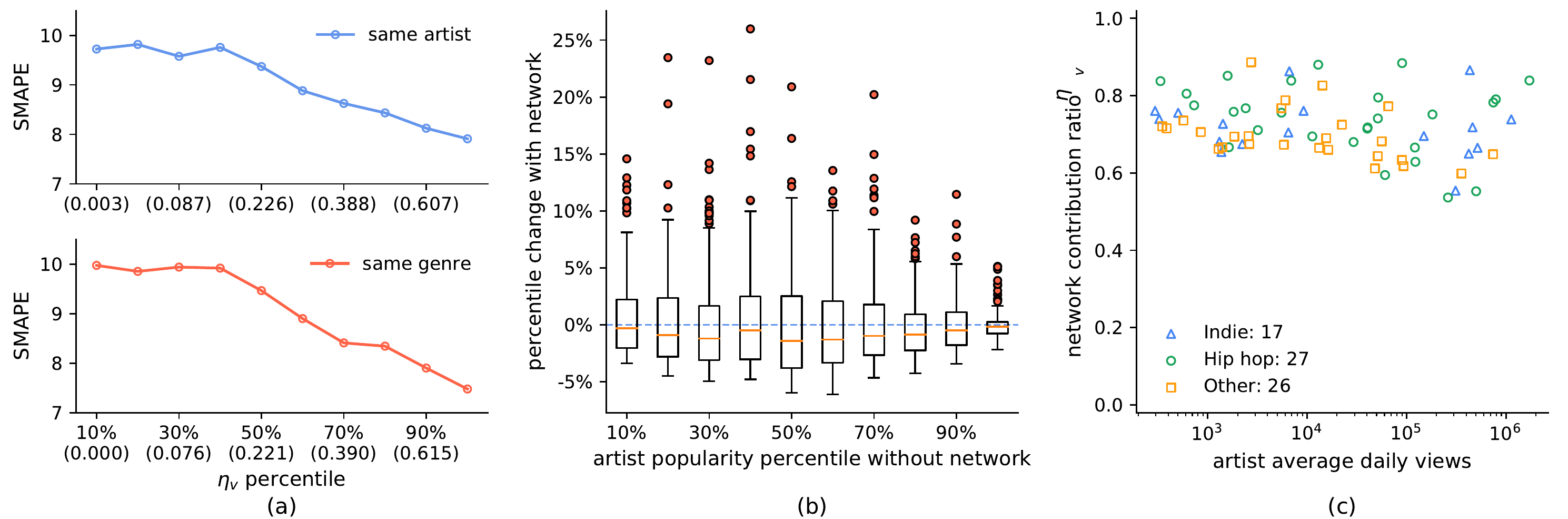}}
	\caption{
	\textbf{(a)} SMAPE as a function of the network contribution $\eta_v$ from videos with the same artist (top) or the same genre (bottom).
	We use the $\eta_v$ percentile as x-axis.
	The numbers within the brackets indicate split values for each percentile, e.g., the right-most dots indicate top 10\% videos with the highest percent of views from similar content, having $\eta_v$ larger than 0.607 for the same artist (top) or 0.615 for the same genre (bottom).
	\textbf{(b)} Boxplot of artists' popularity percentile changes when adding the recommendation network. 
	x-axis: popularity percentile if removing the network; y-axis: popularity percentile change with network.
	The outliers (red circles) denote the artists who gain the most popularity through the network among their cohort. 
	\textbf{(c)} A closer look of artists identified in \textbf{(b)}. A group of Hip hop artists and Indie artists rely more on the recommendation network to become popular.
	}
	\label{fig:model-error}
\end{figure*}

Firstly, we explore the relation between prediction performance and content similarity concerning the artist and music genre.
In \cref{fig:model-error}(a), we compute $\eta_v$ conditioned on that ${(u, v) \in G}$ and that $u$ and $v$ are from the same artist (top) or with the same genre (bottom).
We then slice the x-axis into 20 bins, 5 percentiles apart, based on the artist/genre network contribution ratio.
We compute the mean SMAPEs for the videos in each bin.
Videos that are connected solely by videos from other artists/genres will be placed in the leftmost bin ($\eta_v = 0$).
The plot shows that the SMAPE error decreases with the increasing percentage of views from videos with the same artist or genre.

Secondly, we study the question that which artists are affected most \textit{if} the recommender systems were to be turned off?
\cref{fig:model-error}(b) shows the popularity percentile \textit{change} at the level of artist.
We first compute the network-subtracted views, i.e., subtracting the network contribution $\sum_{t=1}^{\mathbf T} \sum_{(u, v) \in G} \beta_{u, v} \mathbf{y}_{u}[t]$ from the observed views $\sum_{t=1}^{\mathbf T} \mathbf{y}_{v}[t]$.
We then aggregate and compute the popularity percentiles for both observed views and network-subtracted views at the level of artist.
The x-axis plots the artists' popularity percentiles without recommendation network, and y-axis plots the percentile changes when turning on the network.
The range of percentiles stays constant between $[0, 100\%]$, reflecting the concept of finite attention --- one video gains popularity at the expense of others.
The top outliers identify artists who gain much more popularity than their peers with similar popularity due to the recommendation network; whereas the bottom outliers represent artists who lose popularity.
There are 2,340 artists having target videos in the persistent network.
We observe that 1,378 (58.89\%) artists losing a small amount of popularity (less than 5\%) while 948 (40.51\%) gaining.
We notice there is no bottom outlier.
On the contrary, the top outliers show that the network can help some artists massively increase their relative popularity (as high as 26\%, J-Kwon (American rapper) in 4th bin).

We take a closer look at the outliers by scattering them in \cref{fig:model-error}(c).
70 artists gain significant popularity from the recommendation network, implying a better utilization of network effects.
We retrieve the artist genres from the music database MusicBrainz, and we notice two notable groups.
One is the Indie group by matching genre keywords ``indie'', ``alternative'', or ``new wave''.
The top 3 most popular Indie artists are 4 Non Blondes, Hoobastank, and The Police.
The other is the Hip hop group by matching genre keywords ``hip hop'', ``rap'', ``reggae'', or ``r\&b''.
The top 3 most popular Hip hop artists are Mark Ronson, French Montana, and Pharrell Williams.
This finding reveals that the recommender systems can lead users to find niche artists.


\section{Conclusion}
\label{sec:conclusion}

This work presents a large-scale study for online videos on YouTube.
We collect a new dataset that consists of 60,740 \vevo music videos, representing some of the most popular music clips and artists. 
We construct the YouTube recommendation network.
We present measurements on the global component structure and temporal persistence of links. 
A model that leverages the network information for predicting video popularity is proposed, which achieves superior results over other baselines.
It also allows us to estimate the amount of attention flow over each recommendation link.
We derive a metric --- estimated network contribution ratio, and we quantify this ratio at both the entire \vevo network level and individual artist level.
To the best of our knowledge, this is the first work that links the video recommendation network structure to the attention consumption for the videos in it.

\header{Discussion.}
Much progress has been made to algorithmically optimize or increase the attention for individual digital item (from videos to products to connections in social networks), whereas the theory about attention flow among different items is still fairly nascent.
Our data includes a series of network snapshots that are constructed by the platform's recommender systems, and visible to both content producers and consumers. 
We believe that the area of understanding the implications of content recommendation networks has many worthy problems and fruitful applications. 
However, definitions and properties of a recommendation network that is fair and transparent to the content hosting site, producers and consumers remain as open issues. 

\header{Limitation and future work.}
The limitations of this work include: interpretations of importance are directly based on regression weights; some observations may not generalize to other digital items other than the most popular music videos; the prediction does not explore all the potential deep learning architecture and parameter tuning.
Future work includes modeling attention flow that takes into account item rank on the relevant list; connecting aggregate attention with individual click streams; and improving deep neural network models, specifically, three directions for us to exploit.
Firstly, extract additional features, such as audio-visual, artist, and network features.
Secondly, measure the relations between estimated link strength and link properties, such as the diversity and/or novelty of the target video relative to the source video~\cite{ziegler2005improving}.
Lastly, train a shared RNN model on videos with similar dynamics for increasing the volume of training data~\cite{figueiredo2016trendlearner}.

\begin{acks}
We thank Swapnil Mishra and our reviewers for their helpful comments.
This work is supported by Asian Office of Aerospace Research and Development (AOARD) Grant 19IOA078, Australian Research Council Discovery Project DP180101985, and a Google PhD Fellowship.
We also thank NeCTAR Research Cloud for providing computational resources, an Australian research platform supported by the National Collaborative Research Infrastructure Strategy.
\end{acks}

%
\bibliographystyle{ACM-Reference-Format}
\bibliography{networked-popularity-ref}

\end{document}